\documentclass[11pt,a4paper]{article}
\usepackage[utf8]{inputenc}
\usepackage{amsmath}
\usepackage{amsfonts}
\usepackage{amssymb}
\usepackage{amsbsy}

\usepackage{graphicx}
\usepackage[left=3cm,right=3cm,top=2.5cm,bottom=2.5cm]{geometry}
\usepackage{algorithm, algorithmic}
\usepackage{mathrsfs}
\usepackage{hyperref}
\usepackage{tikz}
\usetikzlibrary{shapes,arrows}

\usepackage{multirow}
\usepackage{enumitem}
\usepackage{pdflscape}
\usepackage{bbm}

\newcommand{\E}{\mathbb{E}}
\newcommand{\R}{\mathbb{R}}
\newcommand{\1}[1]{\mathbbm{1}_{\{#1\}}}
\newcommand{\vx}{\boldsymbol{x}}
\newcommand{\vb}{\boldsymbol{b}}
\newcommand{\vbv}{\boldsymbol{b}_{\text{v}}}
\newcommand{\vbb}{\boldsymbol{b}_{\text{b}}}
\newcommand{\bv}{{b}_{\text{v}}}
\newcommand{\bb}{{b}_{\text{b}}}
\newcommand{\mW}{\boldsymbol{W}}
\newcommand{\mWv}{\boldsymbol{W}_{\text{v}}}
\newcommand{\mWb}{\boldsymbol{W}_{\text{b}}}

\newcommand{\nWv}{\mathcal{W}_{\text{v}}}
\newcommand{\nWb}{\mathcal{W}_{\text{b}}}
\newcommand{\dt}{\text{d}t}
\newcommand{\dW}{\text{d}W}
\newcommand{\dS}{\text{d}S}
\newcommand{\dd}[1]{\text{d}#1}
\newcommand{\vc}{C_{\text{v}}}
\newcommand{\vp}{P_{\text{v}}}
\newcommand{\uoc}{C_{\text{u-o}}}
\newcommand{\uic}{C_{\text{u-i}}}
\newcommand{\doc}{C_{\text{d-o}}}
\newcommand{\dic}{C_{\text{d-i}}}
\newcommand{\uop}{P_{\text{u-o}}}
\newcommand{\uip}{P_{\text{u-i}}}
\newcommand{\dop}{P_{\text{d-o}}}
\newcommand{\dip}{P_{\text{d-i}}}
\newcommand{\cd}{C_{\text{d}}}
\newcommand{\pd}{P_{\text{d}}}
\newcommand{\lb}{{\text{m}}}
\newcommand{\ub}{{\text{M}}}

\title{\textbf{Solving barrier options under stochastic volatility using deep learning}}
\date{ }
\author{Weilong Fu\footnote{Department of IEOR, Columbia University, \url{wf2232@columbia.edu}}, Ali Hirsa\footnote{Department of IEOR, Columbia University, \url{ah2347@columbia.edu}}}

\begin{document}
\maketitle

\abstract{We develop an unsupervised deep learning method to solve the barrier options under the Bergomi model. The neural networks serve as the approximate option surfaces and are trained to satisfy the PDE as well as the boundary conditions. Two singular terms are added to the neural networks to deal with the non-smooth and discontinuous payoff at the strike and barrier levels so that the neural networks can replicate the asymptotic behaviors of barrier options at short maturities. After that, vanilla options and barrier options are priced in a single framework. Also, neural networks are employed to deal with the high dimensionality of the function input in the Bergomi model. Once trained, the neural network solution yields fast and accurate option values.}

\providecommand{\keywords}[1]{{\textit{Keywords:}} #1}

\section{Introduction}
Stochastic volatility models are good at replicating the volatility smiles and the correlation between the underlying asset and volatility among the pure diffusion frameworks. Some examples are the Heston model \cite{heston1993closed}, the SABR model \cite{hagan2002managing} and the Bergomi model \cite{bergomi2008smile}. The Bergomi model is more complex since it includes multiple volatility factors and is shown to be better at replicating the term structure of forward variances. However, since the stochastic volatility models define additional dynamics of volatility, option pricing under these models is generally more challenging than that under the models which only consider dynamics of the underlying asset.

Barrier options are path-dependent options whose payoff depends on whether or not the underlying asset has reached the barrier level. They are classified into up/down-and-in/out calls/puts based on the position of the barrier level, its payoff after the barrier level is reached, and the corresponding vanilla option. Traditional methods to price barrier options under the stochastic volatility models include the finite difference method in \cite{chiarella2012evaluation,guardasoni2016fast,kirkby2017unified} and the simulation method in \cite{achtsis2013conditional,cuomo2020sequential}. Also, an analytic approximation for barrier options under stochastic volatility models was proposed in \cite{funahashi2018analytical}.

Aside from traditional methods for option pricing, deep learning methods draw more attention recently.
\begin{itemize}
	\item In supervised deep learning, the neural network is fitted as a function of the option surface or volatility surface w.r.t. all the parameters in the model given labels generated by other pricing methods (see e.g. \cite{beck2018solving, ferguson2018deeply, tugce2019, liu_neural_2019, itkin_deep_2019, bayer2019deep}). The neural network approach is fast in computing prices and volatilities once trained and thus it is a good choice for model calibration. However, in supervised learning, it is pretty costly to generate the training labels by other pricing methods, e.g. finite differences, fast Fourier transform, or simulation.
	\item Here are two main unsupervised approaches:
	\begin{itemize}
	\item In \cite{SIRIGNANO20181339,fu2022unsupervised}, the option prices are solved by PDEs using deep learning. The idea to solve PDEs by deep learning went back to \cite{lee1990neural, lagaris1998artificial, raissi2018deep}. Smooth neural networks are employed as the approximated solution to the PDE and they are trained to match the PDE and boundary conditions. This approach has been applied to vanilla options but not yet to barrier options because common smooth neural networks cannot match the discontinuous boundary conditions or replicate the asymptotic behaviors of barrier options at short maturity. This is also the problem that we need to address in this paper.
	\item In \cite{han2018solving}, the option prices are solved by forward-backward stochastic differential equations. Neural networks are used to approximate the diffusion term in the stochastic differential equations, which is related to the gradient of the solution. Since then, some variants have been applied to the barrier options in \cite{yu2019deep,ganesan2020pricing}.  
	\end{itemize}   
\end{itemize} 

The goal of the paper is to extend the deep learning approach using PDE to the barrier options. In this paper, we propose a pricing method that includes vanilla and barrier options for stochastic volatility models and test it under the Bergomi model. The Bergomi model is a multi-factor stochastic volatility model, which contains more parameters and also a function input. Deep learning is employed to deal with the high dimensionality of the parameter space in the Bergomi model, and is also applicable to the other stochastic models with fewer parameters. 

 In the proposed method, we fit option price surfaces with neural networks. The biggest challenge for a smooth neural network to fit the barrier options is to fit the discontinuous boundary conditions. Thus we propose two singular terms \cite{fu2022unsupervised} and embed them into the neural networks such that the neural networks are not smooth at given points, i.e., the strike and barrier levels at maturity, but are smooth anywhere else. In this way, the networks are able to satisfy the boundary conditions and the PDE at the same time. We train the neural networks with different parameters and the neural networks calculate option values fast after being trained.

The paper is organized as follows. In Section \ref{sec:problem}, we introduce the Bergomi model, the definition of the vanilla and barrier options and the equation groups used for option pricing under the Bergomi model. In Section \ref{sec:roadmap}, we generally introduce the singular terms used for the vanilla and barrier options and the framework of option pricing. In Section \ref{sec:vanilla}, we give the definition of the singular term and the neural network for the vanilla options. We also discuss the boundary conditions of the volatility factors and the loss functions used to train the networks of vanilla options. In Section \ref{sec:barrier}, we give the definition of the singular term, the neural network and the loss functions for the barrier options. In Section \ref{sec:num_exp}, we give the details of numerical experiments, including the piecewise constant function input, the range and distribution of samples for training and the hyperparameters of the neural networks. We show the numerical results from the fitted neural network solutions in terms of the root mean squared error, the relative error and the calculation speed. Section \ref{sec:conclusion} summarizes the paper.

\section{Problem}\label{sec:problem}
In this paper, we focus on solving the barrier options under the Bergomi model, which is a multi-factor stochastic volatility model. It is a general framework proposed by \cite{bergomi2008smile} to capture forward volatility and forward skew risks. The proposed pricing routine is also applicable to other stochastic volatility models after modifications of the boundary conditions of volatility, and of course the basic case of the Black-Merton-Scholes (BMS) model \cite{black_scholes_1973}. In the case of the Bergomi model, we will see how deep learning is employed to deal with the high dimensionality embedded in the model. 
 
\subsection{Bergomi model}\label{subsec:bergomi}
The general $n$-factor Bergomi model is based on the following lognormal dynamics of the forward variances in \cite{bergomi2008smile}
\begin{align*}
	\xi_{t}^{T}=\xi_{0}^{T}\exp\left(\omega\sum_{1\leq i\leq n}w_i e^{-k_i(T-t)}X_t^{(i)}-\frac{\omega^2}{2}\sum_{1\leq i,j\leq n}w_i w_j e^{-(k_i+k_j)(T-t)} \E\left(X_t^{(i)} X_t^{(j)}\right)\right)
\end{align*}
where 
\begin{itemize}
	\item $\xi_{t}^{T},0\leq t\leq T,$ is the process of forward instantaneous variance for date $T$ observed at $t$,
	\item $\xi_{0}^{T},T\geq 0,$ is the initial value of forward variances and is also an input of the model,
	\item $X_t^{(i)},\forall 1\leq i\leq n,$ are OU processes that satisfy $\dd{X}_t^{(i)}=-k_i X_t^{(i)}\dt+\dW_t^{(i)}$ and $X_0^{(i)}=0$,
	\item $w_i,\forall 1\leq i\leq n,$ are positive weights and $\omega$ is a global scaling factor for the volatility of forward variances,
	\item $W_t^{(i)},\forall 1\leq i\leq n,$ are correlated Brownian motions, where $\dW_t^{(i)}\dW_t^{(j)}=\rho_{i,j}\dt$.
\end{itemize}

In \cite{bergomi2008smile}, the author claims that two factors ($n=2$) afford adequate control on the term structure of volatilities of volatilities, and then the multi-factor model is simplified as the two-factor model. Let $0\leq\theta\leq 1$ be a constant and $$\alpha_{\theta}=1/\sqrt{(1-\theta)^2+\theta^2+2\rho_{1,2}\theta(1-\theta)}.$$ The weights in the two-factor model are $w_1=\alpha_{\theta}(1-\theta)$ and $w_2=\alpha_{\theta}\theta$. By introducing the notation 
\begin{align*}
	x_t^T = \alpha_{\theta}\left((1-\theta) e^{-k_1(T-t)}X_t^{(1)}+\theta e^{-k_2(T-t)}X_t^{(2)}\right),
\end{align*}
the dynamics of the forward variances can be simplified as
\begin{align*}
	\xi_{t}^{T}=\xi_{0}^{T}\exp\left(\omega x_t^T-\frac{\omega^2}{2}\text{var}(x_t^T)\right).
\end{align*}

The risk neutral stock price $S_t$ is given by
\begin{align*}
	\dS_t=(r-q)S_t\dt + S_t\sqrt{\xi_t^t}\dW_t^{(S)}
\end{align*}
where $r$ is the risk-free interest rate, $q$ is the dividend rate, $\dW_t^{(S)}\dW_t^{(i)}=\rho_{i}\dt,\forall i=1,2$, and $\xi_t^t$ is given by
\begin{align}
\begin{split}
	\xi_{t}^{t}&=\xi_{0}^{t}\exp\left(\omega x_t^t-\frac{\omega^2}{2}\text{var}(x_t^t)\right)\\
	x_t^t&=\alpha_{\theta}\left((1-\theta)X_t^{(1)}+\theta X_t^{(2)}\right)\\
	\text{var}(x_t^t) &= \alpha_{\theta}^2\,\left((1-\theta)^2 \frac{1-e^{-2k_1 t}}{2k_1} + \theta^2\frac{1-e^{-2k_2 t}}{2k_2} + 2\theta(1-\theta)\rho_{1,2}\frac{1-e^{-(k_1+k_2) t}}{k_1+k_2} \right).	
\end{split}\label{eq:xi_expression}
\end{align}

\subsection{Option pricing}
Suppose $\{S_t\}_{t\geq 0}$ is the stock price process, $s=\ln(S_t)$ is the log-price, $K$ is the strike, $B$ is the barrier level, $t$ is the current time and $T$ is the maturity (expiration) time. Denote the maximum and minimum of the stock price path as 
$$m_t^T=\min_{t\leq \bar{t}\leq T}S_{\bar{t}}\,\,\text{and}\,\,M_t^T=\max_{t\leq \bar{t}\leq T}S_{\bar{t}}.$$
The vanilla/barrier calls/puts are defined as 
\begin{align*}
	V(s,t,x_1,x_2)=e^{-r(T-t)}\E\left(\text{payoff}\,\vert S_t=e^{s},X_t^{(1)}=x_1,X_t^{(2)}=x_2\right)
\end{align*}
where $V$ and payoff are replaced by the corresponding notation and formula in Table \ref{tab:option_def}. 
\begin{table}[h]
	\centering
	\begin{tabular}{lll}\hline
	Option & $V$ &  payoff\\ \hline
vanilla call & $\vc$ & $(S_T-K)^{+}$ \\
vanilla put & $\vp$ & $(K-S_T)^{+}$\\
up-and-out call & $\uoc$ & $(S_T-K)^{+}\1{M_t^T<B}$ \\
up-and-in call & $\uic$ & $(S_T-K)^{+}\1{M_t^T\geq B}$ \\
down-and-out call & $\doc$ & $(S_T-K)^{+}\1{m_t^T>B}$ \\
down-and-in call & $\dic$ & $(S_T-K)^{+}\1{m_t^T\leq B}$ \\
up-and-out put & $\uop$ & $(K-S_T)^{+}\1{M_t^T<B}$ \\
up-and-in put & $\uip$ & $(K-S_T)^{+}\1{M_t^T\geq B}$ \\
down-and-out put & $\dop$ & $(K-S_T)^{+}\1{m_t^T>B}$ \\
down-and-in put & $\dip$ & $(K-S_T)^{+}\1{m_t^T\leq B}$ \\ \hline 
	\end{tabular}
	\caption{Payoffs of vanilla/barrier calls/puts.}
	\label{tab:option_def}
\end{table}

The barrier options satisfy the following in-out parities according to their definitions.
\begin{align}
\begin{cases}
	\uoc(s,t,x_1,x_2)+\uic(s,t,x_1,x_2)=\vc(s,t,x_1,x_2),\forall s,x_1,x_2\in\R,0\leq t\leq T \\
	\doc(s,t,x_1,x_2)+\dic(s,t,x_1,x_2)=\vc(s,t,x_1,x_2),\forall s,x_1,x_2\in\R,0\leq t\leq T\\
	\uop(s,t,x_1,x_2)+\uip(s,t,x_1,x_2)=\vp(s,t,x_1,x_2),\forall s,x_1,x_2\in\R,0\leq t\leq T\\
	\dop(s,t,x_1,x_2)+\dip(s,t,x_1,x_2)=\vp(s,t,x_1,x_2),\forall s,x_1,x_2\in\R,0\leq t\leq T\\
\end{cases}\label{eq:parity}
\end{align}

Our goal is to solve the option values at time $t=0$, i.e., $V(s,0,0,0)$.

\subsection{Equations for option pricing}
Using the Feynman-Kac formula \cite{kac1949distributions}, we can derive the PDE for the two-factor Bergomi model (see Appendix \ref{app:pde}). The option value $V(s,t,x_1,x_2)$ needs to satisfies the following equation in the applicable region for each option:
\begin{align}
\begin{split}
	H(V,s,t,x_1,x_2)=\left(\frac{\partial V}{\partial t} -r V + (r-q-\frac{1}{2}\sigma^2(t,x_1,x_2))\frac{\partial V}{\partial s} \right. &\\
	-k_1 x_1\frac{\partial V}{\partial x_1}-k_2 x_2\frac{\partial V}{\partial x_2} +\frac{1}{2}\sigma^2(t,x_1,x_2)\frac{\partial^2 V}{\partial s^2}+\frac{1}{2}\frac{\partial^2 V}{\partial x_1^2}+\frac{1}{2}\frac{\partial^2 V}{\partial x_2^2} &\\
	\left.+ \rho_1 \sigma(t,x_1,x_2)\frac{\partial^2 V}{\partial s \partial x_1} + \rho_2 \sigma(t,x_1,x_2)\frac{\partial^2 V}{\partial s \partial x_2} + \rho_{1,2}\frac{\partial^2 V}{\partial x_1 \partial x_2}\right) &= 0 	
\end{split}\label{eq:pde_bergomi}
\end{align}
where $\sigma(t,x_1, x_2)$ satisfies $\sigma^2(t,X^{(1)}_t, X^{(2)}_t)=\xi_{t}^{t}$ in Equation \eqref{eq:xi_expression}.

The equation groups including the boundary conditions of the vanilla and knock-in options are listed in Table \ref{tab:option_equation}. The value of the knock-out options can be easily got by the in-out parity in Equation \eqref{eq:parity}. Additionally, each option value $V(s,t,x_1,x_2)$ is continuous during $0\leq t<T$.
\begin{table}[h]
	\centering
	\begin{tabular}{lll}\hline
	Option & Equations\\ \hline
vanilla call & $\begin{cases}
	H(\vc,s,t,x_1,x_2)=0, & \forall s,x_1,x_2\in\R, 0< t< T\\
	\vc(s,T,x_1,x_2) = (e^s-K)^{+}, & \forall s,x_1,x_2\in\R
\end{cases}$ \\
vanilla put & $\begin{cases}
	H(\vp,s,t,x_1,x_2)=0, & \forall s,x_1,x_2\in\R, 0< t< T\\
	\vp(s,T,x_1,x_2) = (K-e^s)^{+}, & \forall s,x_1,x_2\in\R
\end{cases}$\\
up-and-in call & $\begin{cases}
	H(\uic,s,t,x_1,x_2)=0, & \forall s<\ln(B),0<t< T,x_1,x_2\in\R\\
	\uic(s,T,x_1,x_2) = 0, & \forall s<\ln(B),x_1,x_2\in\R\\
	\uic(s,t,x_1,x_2) = \vc(s,t,x_1,x_2), & \forall s\geq\ln(B),0\leq t\leq T,x_1,x_2\in\R\\
\end{cases}$ \\
down-and-in call & $\begin{cases}
	H(\dic,s,t,x_1,x_2)=0, & \forall s>\ln(B),0<t< T,x_1,x_2\in\R\\
	\dic(s,T,x_1,x_2) = 0, & \forall s>\ln(B),x_1,x_2\in\R\\
	\dic(s,t,x_1,x_2) = \vc(s,t,x_1,x_2), & \forall s\leq\ln(B),0\leq t\leq T,x_1,x_2\in\R\\
\end{cases}$\\
up-and-in put & $\begin{cases}
	H(\uip,s,t,x_1,x_2)=0, & \forall s<\ln(B),0<t< T,x_1,x_2\in\R\\
	\uip(s,T,x_1,x_2) = 0, & \forall s<\ln(B),x_1,x_2\in\R\\
	\uip(s,t,x_1,x_2) = \vp(s,t,x_1,x_2), & \forall s\geq\ln(B),0\leq t\leq T,x_1,x_2\in\R\\
\end{cases}$ \\
down-and-in put & $\begin{cases}
	H(\dip,s,t,x_1,x_2)=0, & \forall s>\ln(B),0<t< T,x_1,x_2\in\R\\
	\dip(s,T,x_1,x_2) = 0, & \forall s>\ln(B),x_1,x_2\in\R\\
	\dip(s,t,x_1,x_2) = \vp(s,t,x_1,x_2), & \forall s\leq\ln(B),0\leq t\leq T,x_1,x_2\in\R\\
\end{cases}$\\ \hline 
	\end{tabular}
	\caption{Equations of vanilla/barrier calls/puts.}
	\label{tab:option_equation}
\end{table}

If we solve the option values for $s,x_1,x_2\in \R$ and $0\leq t\leq T$, we also know $V(s,0,0,0)$ as a result.

\subsection{Goal of the paper}
Our goal is to solve the equations in Table \ref{tab:option_equation} using neural networks directly. The option value $V(\vx)$ is treated as a function of not only the variables $s,t,x_1,x_2$, but also all the inputs of the model
\begin{align*}
	\vx=(s,t,x_1,x_2,T,B,r,q,\xi_0^t,\omega,k_1,k_2,\theta,\rho_1,\rho_2,\rho_{1,2}).
\end{align*}
Throughout the paper, the strike $K$ is kept fixed. The function $V(\vx)$ will be approximated by a well-trained neural network. Once the neural network is trained, its output is the option value, and the neural network is able to calculate option values given different parameter sets instantly. Also, no labels of option values from other pricing methods are needed during the training process, so the proposed method is an unsupervised deep learning approach. 

\section{Roadmap}\label{sec:roadmap}
\subsection{Smooth neural network}
The smooth neural networks have been already used to solve PDEs in literature. In \cite{lee1990neural, lagaris1998artificial, raissi2018deep, SIRIGNANO20181339}, the neural network is a function of the space and time variables, while in \cite{fu2022unsupervised}, it is a function of both variables and parameters. The loss of squared residuals of the PDE as well as some boundary conditions is minimized such that the neural network satisfies the equation group.

The building block of the neural networks in this paper is the multi-layer perceptron (MLP). Here we give a quick introduction of the smooth MLP. An MLP is a multi-dimensional function with an input $\vx\in\R^{n_0}$ and an output $V(\vx)\in\R$, where $n_0$ is the length of the input. An MLP with $L$ hidden layers can be constructed by the equations 
\begin{align*}
	\vx^{(0)}&=\vx, \\
	\vx^{(j)}&=g(\mW^{(j-1)}\vx^{(j-1)}+\vb^{(j-1)}), \,\forall 1\leq j\leq L,\\
	V(\vx) &= \mW^{(L)}\vx^{(L)}+b^{(L)},
\end{align*}
where the hidden layers are $\vx^{(j)}\in\R^{n},\forall 1\leq j\leq L$ and the parameters are $\mW^{(0)}\in \mathbb{R}^{n\times n_0}$, $\mW^{(j)}\in \mathbb{R}^{n\times n}$ for $1\leq j\leq L-1$, $\vb^{(j)}\in \mathbb{R}^{n}$ for $0\leq j\leq L-1$, $\mW^{(L)}\in \mathbb{R}^{1\times n}$ and $b^{(L)}\in \mathbb{R}$.
$g$ is the non-linear activation function which is applied element-wise. There are some examples of smooth activation functions in Table \ref{tab:act_fun}. We are going to use SiLU \cite{elfwing2017sigmoidweighted} as the activation function in the neural network since it is empirically shown that it outperforms the other smooth activation functions. Nonetheless, the sigmoid function and the softplus \cite{dugas2000incorporating} function also play important roles in the neural network, which will be covered in the following sections.

\begin{table}[h]
\centering
	\begin{tabular}{cc}\hline
		Function  & Definition\\\hline
		sigmoid & $1/(1+e^{-z})$\\
		SiLU  & $z/(1+e^{-z})$ \\
		softplus  & $\ln(1+e^{z})$\\\hline
	\end{tabular}
	\caption{Examples of smooth activation functions}
	\label{tab:act_fun}
\end{table}

\subsection{Singular terms}
The largest challenge to apply the smooth neural network approach to the barrier options is that their final payoffs at $(s,t)=(\ln(B),T)$ are not continuous. At first glance, we might be able to use the Heaviside function $g(z)=\1{z>0}$ as the activation in the neural network to approximate the discontinuous payoffs. However, the option surface is continuous any time prior to maturity, i.e., for any $t<T$, making the Heaviside function impossible to be used in the neural network. What makes it more challenging is that the solution is discontinuous at one point but continuous anywhere else.

Actually, the discontinuity point $(s,t)=(\ln(B),T)$ is not the only special point. In vanilla options and some barrier options, the point $(s,t)=(\ln(K),T)$ is also a singular point, since their final payoffs are not smooth at this point. A traditional smooth neural network cannot fit well around this point. In \cite{fu2022unsupervised}, a special structure called singular term is used to deal with the non-smoothness around  $(s,t)=(\ln(K),T)$. 

A singular term is a pre-defined function with specific non-smoothness. It is non-smooth (or discontinuous) at maturity but smooth (or continuous) before maturity and that is exactly what we need. Also, they are able to mimic the asymptotic behaviors around the singular point. The input of the singular term consists of trainable components such that the singular term is able to fit the option surface under different parameters. In this paper, we are going to follow the idea of singular terms and propose two singular terms for the two singular points on the option surface $(s,t)=(\ln(K),T)$ and $(s,t)=(\ln(B),T)$, such that we extend the smooth neural network approach to the barrier options.

\subsection{Framework for both vanilla and barrier options}
\begin{figure}[h]
\centering
	\includegraphics[width=0.6\textwidth]{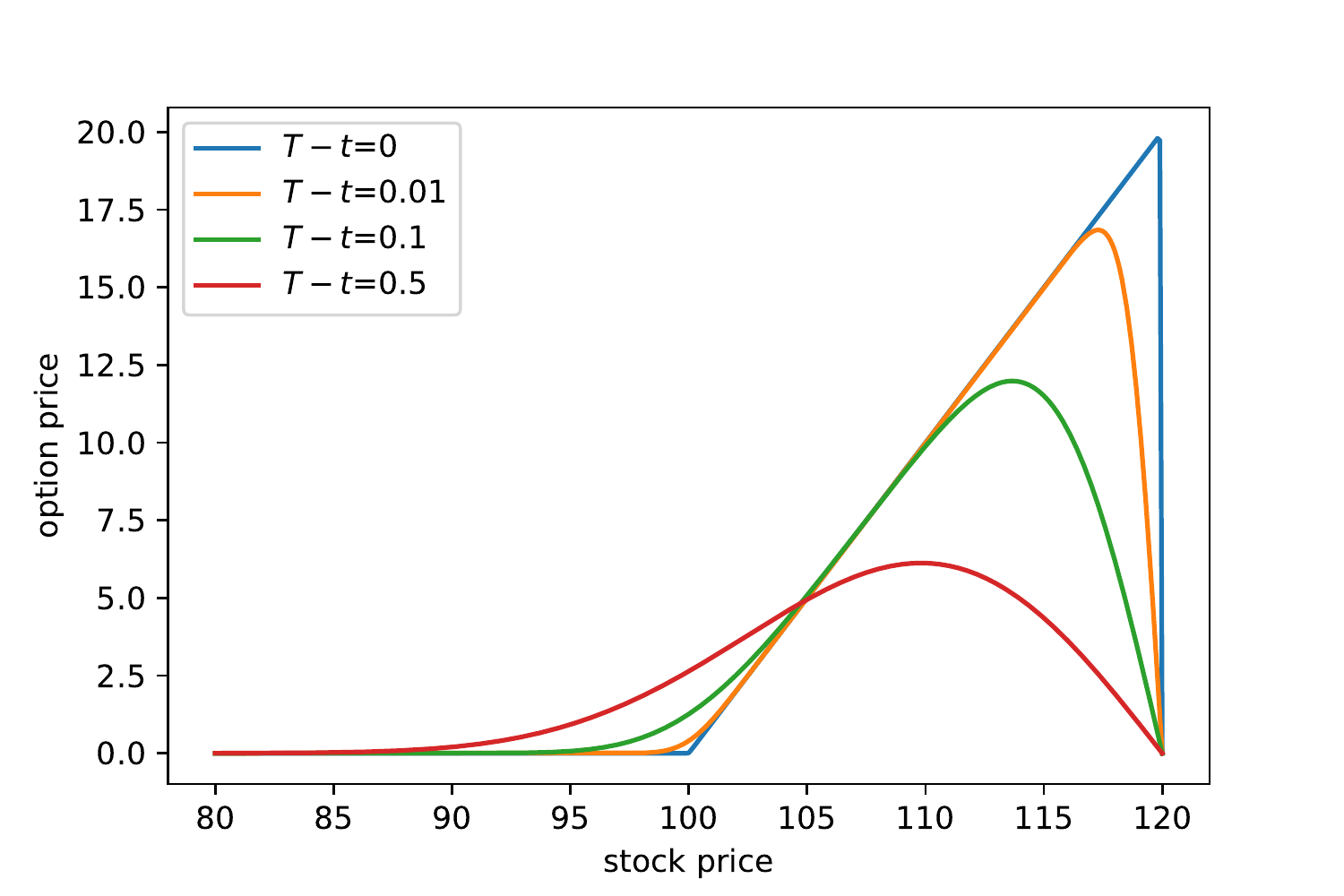}
\caption{Example curves of the up-and-out call when $K=100,B=120,T=0.5,r=q=0$ and $\xi_0^t=0.01$.}
\label{fig:example_uoc}
\end{figure}

We are going to explain how to solve the eight barrier options in a single framework. Take the up-and-out call as an example. Its payoff and the option values prior to maturity are illustrated in Figure \ref{fig:example_uoc}. The option surface of the up-and-out call contains two singular points. So the neural network solution to the up-and-out call needs to contain two singular terms. We have to admit that training the singular term at the barrier level is more challenging than training the one at the strike, since the option surface is continuous at the strike but discontinuous at the barrier level. Thus it is better not to train the two singular terms at the same time. Fortunately, the option surface of all the knock-in options contains just one singular point. We can solve the knock-in options and then the knock-out options are solved by the in-out parity in Equation \eqref{eq:parity} if we also solve the vanilla options.

In the pricing framework, we use six networks to model two vanilla options and eight barrier options: two networks for the vanilla call and put, and four networks for the four knock-in options. We first train the neural networks for vanilla options and then train the networks for knock-in options with the help of the vanilla options. Then each knock-out option is the difference of the corresponding vanilla option and the corresponding knock-in option. Since the up-and-out call and up-and-in call degenerate to 0 and the vanilla call when $B<K$ and the down-and-out put and down-and-in put degenerate to 0 and the vanilla put when $B>K$, we only solve the barrier options in the region where they are non-degenerate. Although we can even use one network for either the vanilla call or put and use the put-call parity to get the other one, we still train them separately using two neural networks.

\section{Vanilla options}\label{sec:vanilla}
\subsection{Singular term for vanilla options}
\begin{figure}[h]
\centering
	\includegraphics[width=0.6\textwidth]{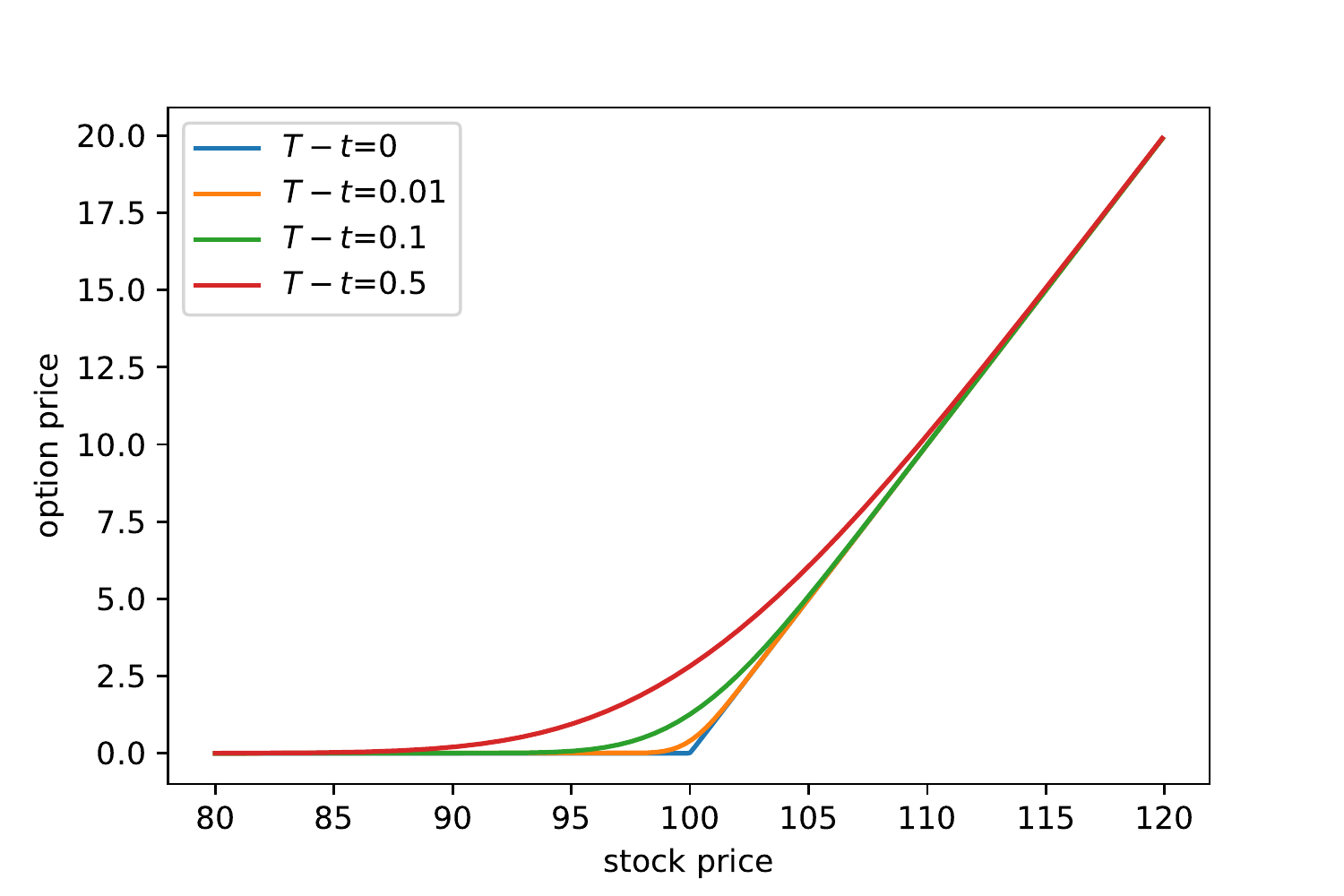}
\caption{Example curves of the vanilla call when $K=100,T=0.5,r=q=0$ and $\xi_0^t=0.01$.}
\label{fig:example_call}
\end{figure}

The option surface of vanilla options is smooth when $t<T$, but not at $(s,t)=(\ln(K),T)$. In Figure \ref{fig:example_call}, we show the call option curve becomes more like a hockey stick at $S=e^s=K$ when $t$ converges to $T$. 
The singular term for vanilla options deals with the singularity around $(s,t)=(\ln(K),T)$. It is modified from the Black-Scholes (BS) formula in Appendix \ref{app:bs_formula}, and it is written as follows:
\begin{align}
\begin{split}
		\alpha_{\text{v}}(\vx)=&\, \eta\, e^{s-q(T-t)}N(\eta( h(\vx)/v(\vx)+  v(\vx)/2))\\
	&-\eta\, K e^{-r(T-t)}N(\eta (h(\vx)/v(\vx) - v(\vx)/2)),\\
	h(\vx) =&\, s - \ln(K) + \beta(\vx)(T-t),\\
	v(\vx) =&\, \gamma(\vx)\sqrt{T-t},
\end{split}\label{eq:singu_v_1}
\end{align}
where $\beta(\vx)$ and $\gamma(\vx)>0$ are both MLPs with an input of $\vx$. The notation $$\eta=\begin{cases}
	+1,&\text{for vanilla and barrier calls}\\
	-1,&\text{for vanilla and barrier puts}
\end{cases}$$ changes the sign according to the option type and will be kept the same hereafter. The function $N(\cdot)$ is the normal CDF and is approximated by 
\begin{align}
	N(z)=\text{sigmoid}\left(2\sqrt{2/\pi}(z+0.044715 z^3)\right)\label{eq:normal_cdf}
\end{align}
in neural networks according to \cite{page1977normal}. Comparing the definition of $\alpha_{\text{v}}(\vx)$ and the BS formula of vanilla options in Appendix \ref{app:bs_formula}, we can find that $r-q$ and $\sigma$ in the BS formula are replaced with $\beta(\vx)$ and $\gamma(\vx)$ in $\alpha_{\text{v}}(\vx)$. The singular term $\alpha_{\text{v}}(\vx)$ satisfies the initial condition of vanilla options
\begin{align*}
	\lim_{t\rightarrow T^{-}} \alpha_{\text{v}}(\vx)=(\eta(e^s-K))^+.
\end{align*}

In \cite{fu2022unsupervised}, a similar singular term is proposed as 
\begin{align*}
	\tilde\alpha_{\text{v}}(\vx)=\text{softplus}\left(\frac{e^{s-q(T-t)}-Ke^{-r(T-t)}+\beta(\vx)(T-t)}{\eta\,\gamma(\vx)\sqrt{T-t}} \right)\gamma(\vx)\sqrt{T-t}
\end{align*}
which is also inspired by the BS formula. The argument inside the softplus function is similar to $h(\vx)/v(\vx)$. The singular term $\tilde\alpha_{\text{v}}(\vx)$ is simpler and also satisfies the initial condition of vanilla options
\begin{align*}
	\lim_{t\rightarrow T^{-}} \tilde\alpha_{\text{v}}(\vx)=(\eta(e^s-K))^+.
\end{align*}
In the Bergomi model, the instant volatility is an exponential function (see Equation \eqref{eq:xi_expression}) and can be very large. The term $\gamma(\vx)$ plays the role of volatility and tends to infinity in some cases. The singular term $\alpha_{\text{v}}(\vx)$ gives the proper limit
\begin{align*}
\lim_{\gamma(\vx)\rightarrow \infty} \alpha_{\text{v}}(\vx)=\begin{cases}
e^{s-q(T-t)}, & \text{for vanilla calls}\\
K e^{-r(T-t)}, & \text{for vanilla puts}
\end{cases}
\end{align*}
in this case but the singular term $\tilde\alpha_{\text{v}}(\vx)$ does not give a proper limit since
\begin{align*}
\lim_{\gamma(\vx)\rightarrow \infty} \tilde\alpha_{\text{v}}(\vx)=\infty.
\end{align*}
So the singular term $\alpha_{\text{v}}(\vx)$ is preferred for the Bergomi model.
                  
\subsection{Dimension reduction}
The Bergomi model is a high-dimensional model not only due to the number of parameters, but also because the model input $\xi_0^t$ is a function. Since the input of the neural network needs to be a vector, we need to consider a family of functions that can be parametrized in a finite-dimensional space, such as step functions or linear functions given fixed nodes. However, the dimension could still be so high such that $\gamma(\vx)$ in Equation \eqref{eq:singu_v_1} needs to learn a very complex volatility surface. Thus we calculate the average of $\xi_0^{t}$ to be 
\begin{align*}
	\bar\sigma_{t}^{T}=\sqrt{\frac{1}{T-t} \int_{t}^T\xi_{0}^{\bar{t}}\, \text{d}{\bar{t}}}
\end{align*} 
and replace the definition of $v(\vx)$ in Equation \eqref{eq:singu_v_1} with 
\begin{align*}
	v(\vx) = \gamma(\vx)\bar\sigma_t^T\sqrt{T-t}.
\end{align*}
The average of $\xi_0^{t}$ lowers the difficulty for $\gamma(\vx)$ to learn the volatility surface.

\subsection{Network structure}
\begin{figure}[h]
\centering
	\begin{tikzpicture}[shorten >=1pt,->,draw=black!50, node distance=2.5 cm]
    \tikzstyle{every pin edge}=[<-,shorten <=1pt]
    \tikzstyle{neuron}=[circle,fill=black!25,minimum size=17pt,inner sep=0pt]
    \tikzstyle{input neuron}=[neuron];
    \tikzstyle{output neuron}=[neuron];
    \tikzstyle{hidden neuron}=[neuron];
    \tikzstyle{annot} = [text width=6em, text centered]

    \foreach \name / \y in {1,...,3}
        \node[input neuron, pin=left:Input] (I-\name) at (0,-\y) {};
    \foreach \name / \y in {1,...,4}
        \path[yshift=0.5cm]
            node[hidden neuron] (H-\name) at (2.5 cm,-\y cm) {};
    \foreach \name / \y in {1,...,4}
        \path[yshift=0.5cm]
            node[hidden neuron] (J-\name) at (5 cm,-\y cm) {};	
	\node[hidden neuron] (S) at (7.5 cm,-1 cm) {$\alpha_{\text{v}}$};          
    \node[hidden neuron] (C) at (7.5 cm,-3 cm) {$m$};       
    \node[output neuron, pin={[pin edge={->}]right:Output}] (O) at (10,-2) {};

    \foreach \source in {1,...,3}
        \foreach \dest in {1,...,4}
            \path (I-\source) edge (H-\dest);    
    \foreach \source in {1,...,4}
        \foreach \dest in {1,...,4}
            \path (H-\source) edge (J-\dest);               
    \foreach \source in {1,...,4}
        \path (J-\source) edge (S); 
    \foreach \source in {1,...,4}
        \path (J-\source) edge (C);
    \path (S) edge (O); 
    \path (C) edge (O); 

    \node[annot,above of=H-1, node distance=1cm] (hl) {Hidden layer $\vx^{(1)}$};
    \node[annot,left of=hl] {Input layer $\vx$};
    \node[annot,right of=hl] (hl2) {Hidden layer $\vx^{(2)}$};
    \node[annot,right of=hl2, node distance=5cm] {Output $w$};
\end{tikzpicture}
\caption{Illustration of the neural network for vanilla options, where $\alpha_{\text{v}}$ is the singular term and $m$ is the smooth term.}
\label{fig:net_vanilla}
\end{figure}
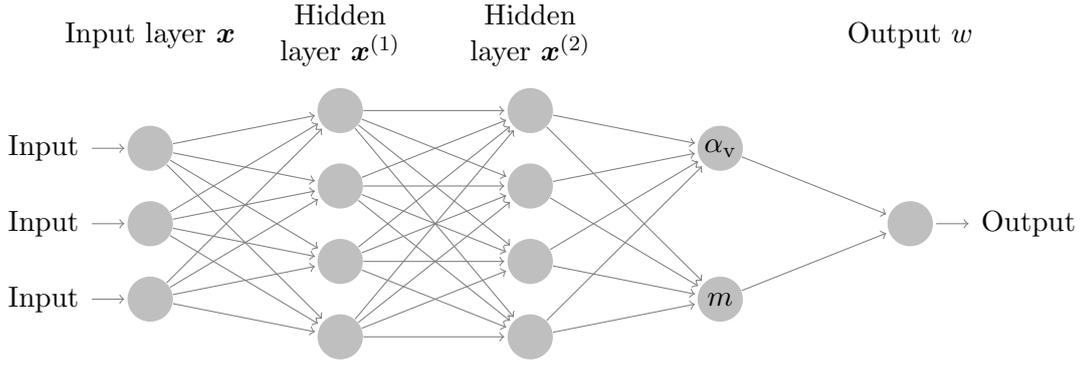

After we introduce the singular term for vanilla options, we give the full expression of the neural network for vanilla options as follows:
\begin{align}
\begin{split}
	\vx^{(0)}=&\, \vx, \\
	\vx^{(j)}=&\, g(\mWv^{(j-1)}\vx^{(j-1)}+\vbv^{(j-1)}), \,\forall 1\leq i\leq L,\\
	\beta(\vx)=&\, \mWv^{(\beta)}\vx^{(L)}+\bv^{(\beta)},\\
	\gamma(\vx)=&\, \text{softplus}\left(\mWv^{(\gamma)}\vx^{(L)}+\bv^{(\gamma)}\right),\\
	h(\vx) =&\, s - \ln(K) + \beta(\vx)(T-t),\\
	v(\vx) =&\, \gamma(\vx)\bar\sigma_t^T\sqrt{T-t},\\
	\alpha_{\text{v}}(\vx)=&\, \eta\, e^{s-q(T-t)}N(\eta( h(\vx)/v(\vx)+  v(\vx)/2))\\
	&-\eta\, K e^{-r(T-t)}N(\eta (h(\vx)/v(\vx) - v(\vx)/2)),\\
	m(\vx) =&\, \sum_{j=0}^{L}\mWv^{(j,V)}\vx^{(j)}+\bv^{(V)},\\
	V(\vx) =&\, m(\vx)+\alpha_{\text{v}}(\vx),	
\end{split}\label{eq:net_vanilla}
\end{align}
where the input layer is $\vx\in\R^{n_0}$ and the hidden layers are $\vx^{(j)}\in\R^{n},\forall 1\leq j\leq L$. $\gamma(\vx)$ is passed through the softplus function to ensure the positivity since it describes the volatility. The singular term $\alpha_{\text{v}}(\vx)$ is built from the last hidden layer $\vx^{(L)}$ and then added to the output. The smooth term $m(\vx)$ has skip connections from all the previous layers $\{\vx^{(j)}\}_{j=0}^{L}$, which stabilize the training process. The output $V(\vx)$ is a sum of the singular term and the smooth term. The dimensions of the neural network parameters are 
\begin{align*}
	\mWv^{(0)}&\in \mathbb{R}^{n\times n_0},\\
	\mWv^{(j)}&\in \mathbb{R}^{n\times n},\forall\,1\leq j\leq L-1,\\
	\mWv^{(j)}&\in \mathbb{R}^{1\times n},\forall\,j=\beta,\gamma,\\
	\mWv^{(0,V)}&\in \mathbb{R}^{1\times n_0},\\
	\mWv^{(j,V)}&\in \mathbb{R}^{1\times n},\forall\,1\leq j\leq L,\\
	\vbv^{(j)}&\in \mathbb{R}^{n},\forall 0\leq j\leq L-1,\\
	\bv^{(j)}&\in \mathbb{R},\forall\,j=\beta,\gamma,V.
\end{align*} 
The overall structure is an MLP with $L$ layers of width $n$, and with a singular term added to the output. A graph of the neural network with $L=2$ is illustrated in Figure \ref{fig:net_vanilla} if we omit the skip connections.

\subsection{Boundary conditions of volatility}\label{subsec:bound_volatility}
Since the PDE contains derivatives w.r.t. the volatility factors $x_1$ and $x_2$, we need to add boundary conditions for them. We need to anchor the solution on the boundary of $x_1$ and $x_2$, otherwise the solution would be far from the true value on the boundary and the solution in the interior would also be inaccurate even if the PDE is satisfied in the interior. 

For the Heston model \cite{heston1993closed}, which is also a stochastic volatility model, the dynamics of the stock price and volatility are 
\begin{align*}
	\dS_t&=r S_t\dt + S\sqrt{V_t}\dW_t^{(S)},\\
	\dd{V}_t&=k(\theta-V_t)\dt + \sigma \sqrt{V_t}\dW_t^{(V)},
\end{align*}
where $V_t$ is the variance process, $k,\theta$ and $\sigma$ are positive constants and $W_t^{(S)}$ and $W_t^{(V)}$ are correlated Brownian motions. Let $C_{\text{H}}(S,v,t)$ be the value of vanilla calls with stock price $S$ and instant volatility $\sqrt{V_t}$ at time $t$ in the Heston model. The theoretical boundary conditions for vanilla calls at $V_t=0$ and $V_t=\infty$ proposed in \cite{heston1993closed} are $$rS\frac{\partial C_{\text{H}}}{\partial S}(S,0,t)+\kappa\theta\frac{\partial C_{\text{H}}}{\partial v}(S,0,t)-r C_{\text{H}}(S,0,t)+\frac{\partial C_{\text{H}}}{\partial t}(S,0,t)=0$$ and 
$$C_{\text{H}}(S,\infty,t)=S.$$
However, this kind of boundary conditions does not work well in practice for the neural network approach. Although we know a vanilla call with infinity volatility converges to the stock price $S$, it is hard to know how large could be considered as `infinity' in the numerical routine. In the Bergomi model, we can get similar results for $x_1=\pm\infty$ and $x_2=\pm\infty$, but the question remains how large `infinity' is.

We need a better estimate of $V(\vx)$ when $x_1$ and $x_2$ are far from 0. Recall that $X_t^{(1)}$ is defined by
\begin{align*}
	\dd{X}_t^{(1)} &= -k_1 {X}_t^{(1)} \dt + \dW_t^{(1)}.
\end{align*}
If ${X}_t^{(1)}=x_1$ and $x_1$ is far from 0, the drift term $-k_1 {X}_t^{(1)} \dt$ dominates in the dynamic. We consider 
\begin{align*}
	\dd{\tilde{X}}_t^{(1)}= -k_1 \tilde{X}_t^{(1)} \dt
\end{align*}
and $$\tilde{X}_u^{(1)}=x_1 e^{-k_1(u-t)},\forall u\geq t$$ is a deterministic function. We also let $$\tilde{X}_u^{(2)}=x_2 e^{-k_2(u-t)},\forall u\geq t.$$ If we replace ${X}_u^{(1)}$ and ${X}_u^{(2)}$ with $\tilde{X}_u^{(1)}$ and $\tilde{X}_u^{(2)}$, then $\xi_u^u,\forall u\geq t$ becomes a deterministic function according to its definition in Equation \eqref{eq:xi_expression}, and the Bergomi model degenerates to the BMS model, where the volatility rate is $$\sqrt{\frac{1}{T-t}\int_{t}^T\xi_u^u \dd{u}}.$$
In this way, we do not require $x_1$ and $x_2$ to be infinity in the boundary condition. They are required to be far from 0 such that we can omit the drift terms in the dynamics of the OU processes. A suitable choice of $x_1$ and $x_2$ could be the quantiles of the limiting distribution of the OU processes, as we do in Section \ref{subsec:range}.

We estimate vanilla options under the Bergomi model when $x_1$ and $x_2$ are far from 0 by vanilla options under the BMS model. Although we have to admit that there are still some errors in the estimation since we do not consider the correlation $\rho_1,\rho_2$ and $\rho_{1,2}$, it is much better than the boundary condition of $C_{\text{H}}(S,\infty,t)=S$ for numerical use.

\subsection{Loss function}
In this part we still write the option value $V(\vx)$ as $V(s,t,x_1,x_2)$ to emphasize the different variables in the boundary conditions. The other parameters are omitted in notations since they will be kept the same in the boundary conditions, but we still need to keep in mind that $V$ is a function of $\vx$. Let $\tilde{V}(s,t,x_1,x_2)$ be the estimate by the BMS model in Section \ref{subsec:bound_volatility} when we replace ${X}_u^{(1)}$ and ${X}_u^{(2)}$ with $\tilde{X}_u^{(1)}$ and $\tilde{X}_u^{(2)}$. Let $s_{\lb}$ and $s_{\ub}$ be the lower and upper boundaries for the variable $s$. Let $x_{j,\lb}(\vx)$ and $x_{j,\ub}(\vx)$ be the lower and upper boundaries for the variable $x_j,\forall j=1,2$. Note that $s_{\lb}$ and $s_{\ub}$ are constants while $x_{j,\lb}(\vx)$ and $x_{j,\ub}(\vx)$ are functions depending on the other parameters in $\vx$. In Table \ref{tab:bound_vanilla}, we list the boundary conditions for vanilla calls and puts. We do not calculate boundary conditions for $x_1$ and $x_2$ separately, since we need both $x_1$ and $x_2$ to be far from 0 so that we can use the estimate $\tilde{V}(s,t,x_1,x_2)$ as the boundary condition.
\begin{table}[h]
\centering
\begin{tabular}{lll}\hline
Boundary value &  Vanilla call & Vanilla put \\ \hline
$V(s,T,x_1,x_2)$ & $(e^{s}-K)^{+}$ & $(K-e^{s})^{+}$\\
$V(s_{\lb},t,x_1,x_2)$ & 0 & $K e^{-r(T-t)}-e^{s_{\lb}-q(T-t)}$\\
$V(s_{\ub},t,x_1,x_2)$ & $e^{s_{\ub}-q(T-t)}-K e^{-r(T-t)}$& 0\\
$V(s,t,x_{1,\lb}(\vx),x_{2,\lb}(\vx))$ & $\tilde{V}(s,t,x_{1,\lb}(\vx),x_{2,\lb}(\vx))$ & $\tilde{V}(s,t,x_{1,\lb}(\vx),x_{2,\lb}(\vx))$ \\
$V(s,t,x_{1,\ub}(\vx),x_{2,\ub}(\vx))$ & $\tilde{V}(s,t,x_{1,\ub}(\vx),x_{2,\ub}(\vx))$ & $\tilde{V}(s,t,x_{1,\ub}(\vx),x_{2,\ub}(\vx))$ \\ \hline 
	\end{tabular}
	\caption{Boundary conditions of vanilla calls and puts.}
	\label{tab:bound_vanilla}
\end{table}

Let $V(\vx)$ be the neural network defined in Equation \eqref{eq:net_vanilla} with parameters 
\begin{align} 
	\nWv=\left\{ \begin{array}{l}
	\mWv^{(j)},\vbv^{(j)},\forall 0\leq j\leq L-1\,\text{or}\,j=\beta,\gamma\\
	\mWv^{(j,V)},\forall 0\leq j\leq L\,\text{and}\,\bv^{(V)}
	\end{array} \right\}.
\end{align}
Given a sample $\vx$, the loss function for vanilla puts is defined as  
\begin{align*}
\begin{split}
L_{\vp}(\nWv;\vx) =& \left(H(V,s,t,x_1,x_2)\right)^2 + \left(V(s,T,x_1,x_2)-(K-e^{s})^{+}\right)^2 \\
&+ \left(V(s_{\lb},t,x_1,x_2)-K e^{-r(T-t)}-e^{s_{\lb}-q(T-t)}\right)^2 + \left(V(s_{\ub},t,x_1,x_2)\right)^2 \\
&+ \lambda_1\left(V(s,t,x_{1,\lb}(\vx),x_{2,\lb}(\vx))-\tilde{V}(s,t,x_{1,\lb}(\vx),x_{2,\lb}(\vx))\right)^2 \\
&+ \lambda_1\left(V(s,t,x_{1,\ub}(\vx),x_{2,\ub}(\vx))-\tilde{V}(s,t,x_{1,\ub}(\vx),x_{2,\ub}(\vx))\right)^2 
\end{split}
\end{align*} 
where $H(V,s,t,x_1,x_2)$ is defined in Equation \eqref{eq:pde_bergomi}. $\lambda_1$ is constant with the default value $\lambda_1=0.01$, which means we allow some errors in the boundary conditions for $x_1$ and $x_2$. 

The loss for vanilla calls is a little different since we need to compensate for the large values and derivatives when $s$ is near the upper boundary $s_{\ub}$ such that they will not dominate the loss function, which is also used in \cite{fu2022unsupervised}. The weight is defined as $$\phi(s)=\min(1,4K^2\exp(-2s))$$ 
since the values and derivatives of vanilla calls grow at the rate of $\exp(s)$. Then the loss function for vanilla calls is defined as 
\begin{align*}
\begin{split}
L_{\vc}(\nWv;\vx) =& \,\,\phi(s)\left(H(V,s,t,x_1,x_2)\right)^2 + \phi(s)\left(V(s,T,x_1,x_2)-(e^{s}-K)^{+}\right)^2 \\
&+ \left(V(s_{\lb},t,x_1,x_2)\right)^2 + \phi(s_{\ub})\left(V(s_{\ub},t,x_1,x_2)-e^{s_{\ub}-q(T-t)}-K e^{-r(T-t)}\right)^2 \\
&+ \lambda_1\phi(s)\left(V(s,t,x_{1,\lb}(\vx),x_{2,\lb}(\vx))-\tilde{V}(s,t,x_{1,\lb}(\vx),x_{2,\lb}(\vx))\right)^2 \\
&+ \lambda_1\phi(s)\left(V(s,t,x_{1,\ub}(\vx),x_{2,\ub}(\vx))-\tilde{V}(s,t,x_{1,\ub}(\vx),x_{2,\ub}(\vx))\right)^2. 
\end{split}
\end{align*}

The losses $L_{\vc}$ and $L_{\vp}$ are defined on a single sample $\vx$. Given multiple samples $\vx_{(1)},\vx_{(2)},\dots,\vx_{(n)}$, the total loss is an average of the individual losses, i.e., 
\begin{align*}
	L_{\text{v,avg}}\left(\nWv;\{\vx_{(j)}\}_{j=1}^n\right)=\begin{cases}
		\frac{1}{n} \sum_{j=1}^{n} L_{\vp}(\nWv;\vx_{(j)}),&\text{for vanilla puts,} \\
		\frac{1}{n} \sum_{j=1}^{n} L_{\vc}(\nWv;\vx_{(j)}),&\text{for vanilla calls.} 
	\end{cases} 
\end{align*}
We minimize the loss function w.r.t. $\nWv$ such that the network $V(\vx)$ approximates the true value of vanilla options.

\section{Barrier options}\label{sec:barrier}
\subsection{Singular term for barrier options}
The singular term for vanilla options is to deal with the singularity around $(s,t)=(\ln(K),T)$. Although the option surface is not smooth around $(s,t)=(\ln(K),T)$, it is continuous. A smooth neural network without the singular term is still able to fit the entire vanilla option surface with small errors, except that it cannot completely meet the initial condition. The singular term is an improvement of the neural network but not a requirement.

\begin{figure}[h]
\centering
\begin{tabular}{cc}
	\includegraphics[width=0.48\textwidth]{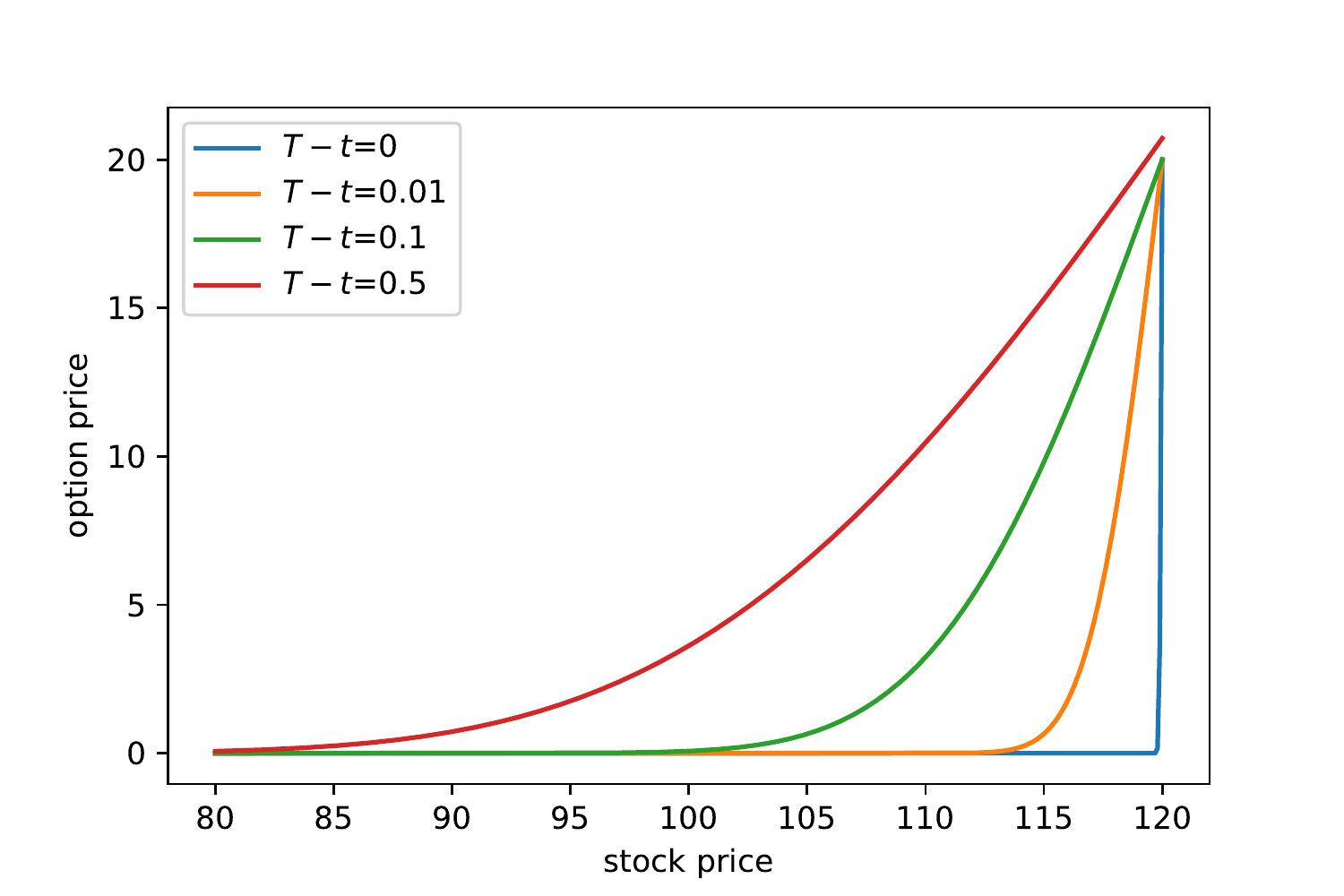}&
	\includegraphics[width=0.48\textwidth]{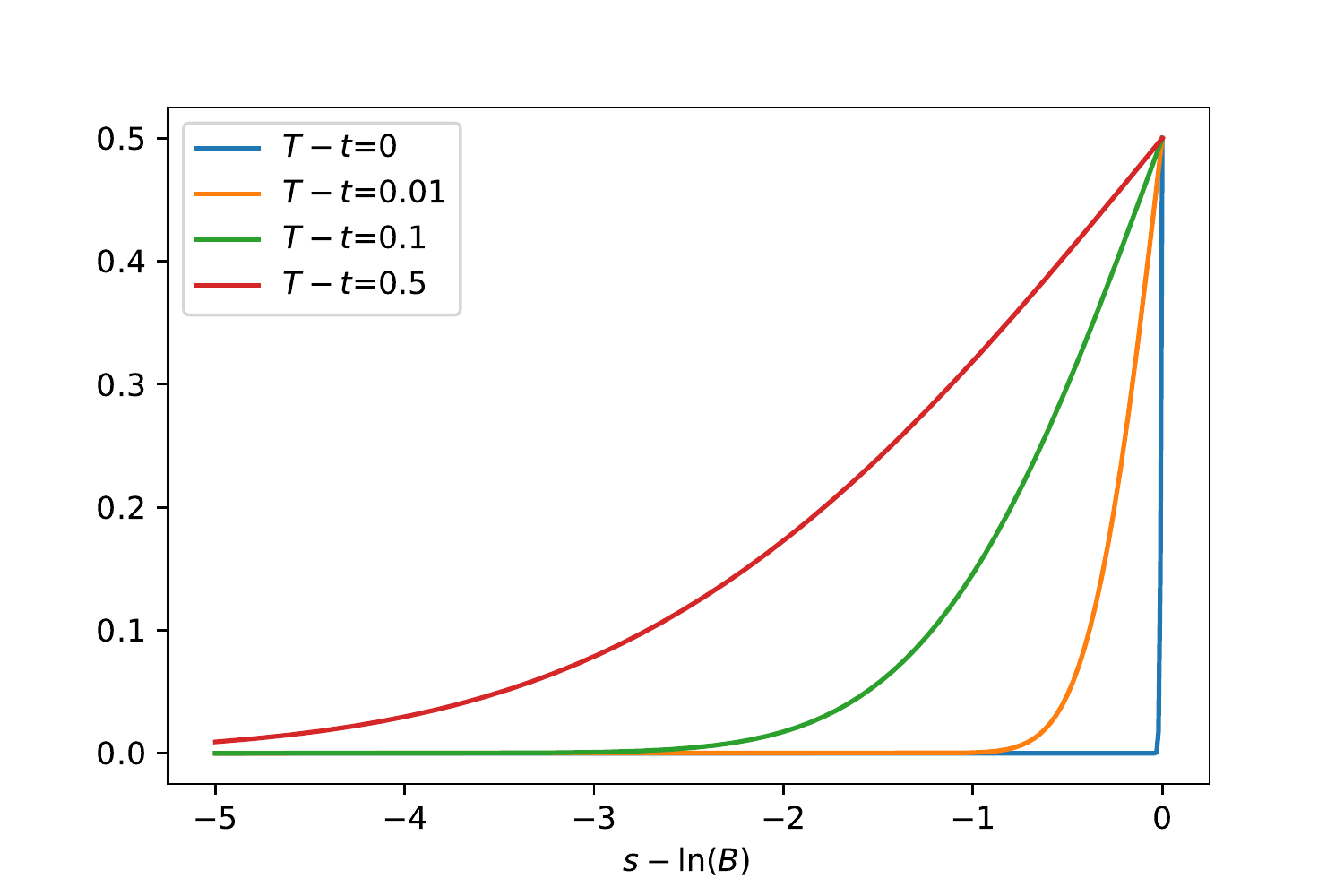}\\
	(a) & (b)\\
\end{tabular}
\caption{(a) Example curves of the up-and-in call when $K=100,B=120,T=0.5,r=q=0$ and $\xi_0^t=0.04$. (b) Example curves of the singular term $F_1(\beta(\vx),\gamma(\vx),\vx)$ when $r-q+\beta(\vx)=0$ and $\gamma(\vx)\bar\sigma_t^T=3$.}
\label{fig:example_uic}
\end{figure}

However, the case is different for the barrier options. In Figure \ref{fig:example_uic} (a), we show the curves of the up-and-in call. As $t$ approaches $T$, the curve becomes more and more vertical near $S=e^s=B$. The option surface is not continuous at $(s,t)=(\ln(B),T)$ and cannot be fitted by a continuous smooth neural network. The optimization routine would fail since the boundary conditions cannot be fitted. 
Thus it is necessary to add the singular term for barrier options to the smooth neural network to overcome this problem:
\begin{align*}
	\alpha_{\text{b}}(\vx)=&\, F_1(\beta(\vx),\gamma(\vx),\vx)\\
	=&\, N(\zeta\,h_{B}(\vx)/v(\vx)),\\
	h_{B}(\vx) =&\, s - \ln(B) + (r-q+\beta(\vx))(T-t),\\
	v(\vx) =&\, \gamma(\vx)\bar\sigma_t^T\sqrt{T-t},
\end{align*}
where $\beta(\vx)$ and $\gamma(\vx)>0$ are MLPs with an input of $\vx$. The notation $\zeta$ is defined as $$\zeta=\begin{cases}
	+1,&\text{for up-and-in options,}\\
	-1,&\text{for down-and-in options.}
\end{cases}$$ 
The normal CDF is approximated by Equation \eqref{eq:normal_cdf}. The singular term is designed such that 
\begin{align*}
	\lim_{t\rightarrow T^{-}} F_1(\beta(\vx),\gamma(\vx),\vx)=\1{\zeta(s-\ln(B))>0}.
\end{align*}
It is a Heaviside function at maturity but is smooth before maturity. In Figure \ref{fig:example_uic} (b), the singular term is similar to the option curves in the region $s<\ln(B)$ when $t$ converges to $T$.

\begin{figure}[h]
\centering
	\includegraphics[width=0.6\textwidth]{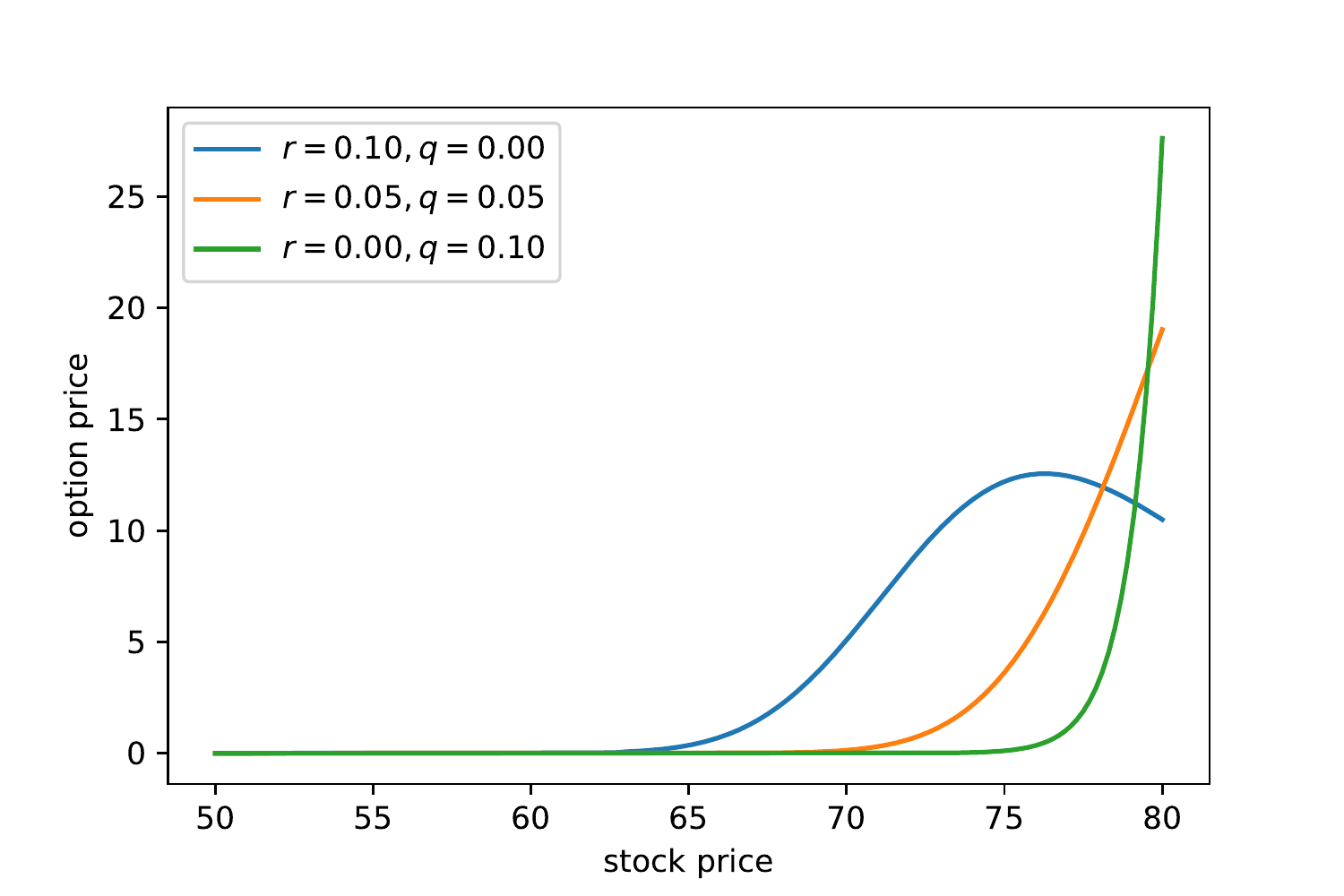}
\caption{Example curves of the up-and-in put when $K=100,B=80,T=1,t=0$ and $\xi_0^t=0.0025$.}
\label{fig:example_uip}
\end{figure}

The singular term $F_1(\beta(\vx),\gamma(\vx),\vx)$ is able to replicate the discontinuity around $(s,t)=(\ln(B),T)$. However, we need to pay special attention to the cases of up-and-in puts and down-and-in calls since their curves are not necessarily monotone w.r.t. the stock price. This is more obvious when volatility is small and the difference between $r$ and $q$ is large. In Figure \ref{fig:example_uip}, we show the curves of the up-and-in put. The curves of up-and-in puts are increasing when $r\leq q$, while the curve is not monotone when $r$ is much larger than $q$. This phenomenon increases the difficulty of fitting at longer maturities since the singular term $F_1(\beta(\vx),\gamma(\vx),\vx)$ is always monotone. 
Once again, we think of the BS formula for the barrier options, which is summarized in Appendix \ref{app:bs_formula}, and propose the following singular term for up-and-in puts and down-and-in calls:
\begin{align*}
\alpha_{\text{b}}(\vx)=&\, F_2(\beta(\vx),\gamma(\vx),\vx)\\
=&\, F_{2,1}(\beta(\vx),\gamma(\vx),\vx) \\
&+ F_{2,2}(\beta(\vx),\gamma(\vx),\vx)\exp((s - \ln(B))(1-2(r-q)/ (\bar\sigma_t^T)^2))
\end{align*}
where the two components are
\begin{align*}
F_{2,1}(\beta(\vx),\gamma(\vx),\vx)= \begin{cases}
\begin{array}{l}
\eta\, e^{s-q(T-t)}N(\eta( h_{K}(\vx)/v(\vx)+  v(\vx)/2))\\
-\eta\, K e^{-r(T-t)}N(\eta (h_{K}(\vx)/v(\vx) - v(\vx)/2))\\
-\eta\, e^{s-q(T-t)}N(\eta( h_{B}(\vx)/v(\vx)+  v(\vx)/2))\\
+\eta\, K e^{-r(T-t)}N(\eta (h_{B}(\vx)/v(\vx) - v(\vx)/2)),
\end{array} & \text{if}\,{\eta(K-B)<0},\\
\,0, &\text{else},
\end{cases}
\end{align*}
and
\begin{align*}
F_{2,2}(\beta(\vx),\gamma(\vx),\vx)=&\, \eta\, B^2e^{-s-q(T-t)}N(\eta( \tilde{h}(\vx)/v(\vx)+  v(\vx)/2))\\
&-\eta\, K e^{-r(T-t)}N(\eta (\tilde{h}(\vx)/v(\vx) - v(\vx)/2)),
\end{align*}
and the elements in the normal CDF are
\begin{align*}
h_{B}(\vx) &= s - \ln(B) + (r-q+\beta(\vx))(T-t),\\
h_{K}(\vx) &= s - \ln(K) + (r-q+\beta(\vx))(T-t),\\
\tilde{h}(\vx) &= \begin{cases}
	2\ln(B)-s -\ln(K) + (r-q+\beta(\vx))(T-t),&\text{if}\,\eta(K-B)\geq 0, \\
	\ln(B)-s + (r-q+\beta(\vx))(T-t),&\text{else},
\end{cases} \\
v(\vx) &= \gamma(\vx)\bar\sigma_t^T\sqrt{T-t}.
\end{align*}
$\beta(x)$ and $\gamma(\vx)>0$ are still two MLPs. The singular term is actually modified from the BS formula for up-and-in puts and down-and-in calls. It is easy to see the singular term replaces $r-q$ and $\sigma$ in the normal CDF with $r-q+\beta(x)$ and $\gamma(\vx)\bar\sigma_t^T$ respectively. While we should be able to modify the BS formula for up-and-in calls and down-and-in puts to get a singular term, $F_1(\beta(\vx),\gamma(\vx),\vx)$ is capable of this job and is beneficial for its simplicity and numerical stability.       
           
\subsection{Network structure}
\begin{figure}[h]
\centering
	\begin{tikzpicture}[shorten >=1pt,->,draw=black!50, node distance=2.5 cm]
    \tikzstyle{every pin edge}=[<-,shorten <=1pt]
    \tikzstyle{neuron}=[circle,fill=black!25,minimum size=17pt,inner sep=0pt]
    \tikzstyle{input neuron}=[neuron];
    \tikzstyle{output neuron}=[neuron];
    \tikzstyle{hidden neuron}=[neuron];
    \tikzstyle{annot} = [text width=6em, text centered]

    \foreach \name / \y in {1,...,3}
        \node[input neuron, pin=left:Input] (I-\name) at (0,-\y) {};
    \foreach \name / \y in {1,...,4}
        \path[yshift=0.5cm]
            node[hidden neuron] (H-\name) at (2.5 cm,-\y cm) {};
    \foreach \name / \y in {1,...,4}
        \path[yshift=0.5cm]
            node[hidden neuron] (J-\name) at (5 cm,-\y cm) {};
	\path[yshift=0.5cm]
            node[hidden neuron] (S) at (3.75 cm,0 cm) {$\alpha_{\text{b}}$};       
    \node[output neuron, pin={[pin edge={->}]right:Output}] (O) at (7.5,-2) {};

    \foreach \source in {1,...,3}
        \foreach \dest in {1,...,4}
            \path (I-\source) edge (H-\dest);   
    \foreach \source in {1,...,4}
        \foreach \dest in {1,...,4}
            \path (H-\source) edge (J-\dest);               
    \foreach \source in {1,...,4}
        \path (J-\source) edge (O);   
    \foreach \source in {1,...,4}
        \path (H-\source) edge (S); 
	\foreach \source in {1,...,4}
        \path (S) edge (J-\source);     

    \node[annot,above of=H-1, node distance=2cm] (hl) {Hidden layer $\vx^{(1)}$};
    \node[annot,left of=hl] {Input layer $\vx$};
    \node[annot,right of=hl] (hl2) {Hidden layer $\vx^{(2)}$};
    \node[annot,right of=hl2] {Output $V(\vx)$};
\end{tikzpicture}
\caption{Illustration of the neural network for knock-in options, where $\alpha_{\text{b}}$ is the singular term.}
\label{fig:net_barrier}
\end{figure}
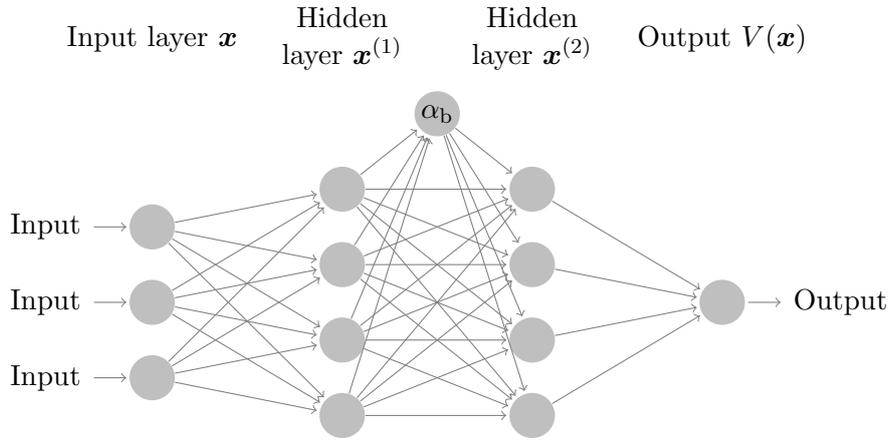

After introducing the singular term for barrier options, we define the neural networks for knock-in options as follows:
\begin{align}
\begin{split}
	\vx^{(0)}&=\vx, \\
	\vx^{(j)}&=g(\mWb^{(j-1)}\vx^{(j-1)}+\vbb^{(j-1)}), \,\forall 1\leq i\leq L_1,\\
	\beta(\vx)&=\mWb^{\beta}\vx^{(L_1)}+\bb^{\beta},\\
	\gamma(\vx)&=\text{softplus}\left(\mWb^{\gamma}\vx^{(L_1)}+\bb^{\gamma}\right),\\
	\alpha_{\text{b}}(\vx)&=\begin{cases}
		F_1(\beta(\vx),\gamma(\vx),\vx),&\text{for up-and-in calls and down-and-in puts,}\\
		F_2(\beta(\vx),\gamma(\vx),\vx),&\text{for up-and-in puts and down-and-in calls,}
	\end{cases} \\
	\tilde{\vx}^{(L_1)} &= \text{concatenate}({\vx}^{(L_1)}, \alpha_{\text{b}}(\vx)),\\
	\vx^{(L_1+1)}&=g\left(\mWb^{(L_1)}\tilde{{\vx}}^{(L_1)}+\vbb^{(L_1)}\right),\\
	\vx^{(j)}&=g\left(\mWb^{(j-1)}\vx^{(j-1)}+\vbb^{(j-1)}\right), \,\forall L_1+1< j\leq L_1+L_2,\\
	V(\vx) &= \mWb^{(L_1+L_2)}\vx^{(L_1+L_2)}+\bb^{(L_1+L_2)},
\end{split}\label{eq:net_barrier}
\end{align}
where the input layer $\vx$ is of size $n_0$ and the hidden layers are $\vx^{(j)}\in\R^{n},\forall 1\leq j\leq L_1+L_2$. The singular term $\alpha_{\text{b}}(\vx)$ is built from the middle hidden layer $\vx^{(L_1)}$ and then combined with $\vx^{(L_1)}$ to be fed to the next hidden layer. The singular term has to be embedded in the middle since we need the neural network to figure out how to combine the singular term and the continuous part by itself. The dimensions of the neural network parameters are 
\begin{align*}
	\mWb^{(0)}&\in \mathbb{R}^{n\times n_0},\\
	\mWb^{(j)}&\in \mathbb{R}^{n\times n},\forall\,1\leq j\leq L_1-1,L_1+1\leq j\leq L_1+L_2-1,\\
	\mWb^{(j)}&\in \mathbb{R}^{1\times n},\forall\,j=\beta,\gamma,L_1+L_2,\\
	\mWb^{(L_1)}&\in \mathbb{R}^{n\times (n+1)}\\
	\vbb^{(j)}&\in \mathbb{R}^{n},\forall 0\leq j\leq L_1+L_2-1,\\
	\bb^{(j)}&\in \mathbb{R},\forall\,j=\beta,\gamma,L_1+L_2.
\end{align*} 
The overall structure is an MLP with $L_1+L_2$ layers of width $n$, and with a singular term embedded in the middle. A graph of the neural network with $L_1=L_2=1$ is illustrated in Figure \ref{fig:net_barrier}.

\subsection{Loss function}
We now go over the boundary conditions and loss functions for barrier options. We only apply the boundary conditions for $s$ at $s_{\lb}$, $\ln(B)$ or $s_{\ub}$ and do not apply the boundary conditions for $x_1$ and $x_2$ for the following two reasons. First, the BMS model cannot serve as an estimate since the barrier options are path-dependent and the instant volatility in the Bergomi model changes fast. Second, the vanilla option value in the boundary conditions for $s$ is already a reference of the barrier options when $x_1$ and $x_2$ are far from 0. The boundary conditions for knock-in options are listed in Table \ref{tab:bound_barrier}. In this part we still only use the arguments $s,t,x_1,x_2$ and omit the other parameters in notations.
\begin{table}[h]
\centering
\begin{tabular}{lllll}\hline
Boundary condition &  Initial & Lower & Middle & Upper \\ 
Options & ($t=T$) & ($s=s_{\lb}$) & ($s=\ln(B)$) & ($s=s_{\ub}$) \\ \hline
up-and-in call & 0 & 0 & $\vc(\ln(B),t,x_1,x_2)$ & N/A \\
up-and-in put & 0 & 0 & $\vp(\ln(B),t,x_1,x_2)$ & N/A \\
down-and-in call & 0 & N/A & $\vc(\ln(B),t,x_1,x_2)$ & 0\\
down-and-in put & 0 & N/A & $\vp(\ln(B),t,x_1,x_2)$ & 0 \\ \hline 
	\end{tabular}
	\caption{Boundary conditions of knock-in options. `N/A' means the boundary condition is not applicable for certain barrier options since the boundary is on the other side of the barrier level compared with the stock price.}
	\label{tab:bound_barrier}
\end{table}

Let $V(\vx)$ be the neural network defined in Equation \eqref{eq:net_barrier} with parameters 
\begin{align*} 
	\nWb=\left\{\mWb^{(j)},\vbb^{(j)},\forall 0\leq j\leq L_1+L_2\,\text{or}\,j=\beta,\gamma \right\}.
\end{align*}
Given a sample $\vx$, the loss functions for up-and-in options are defined as  
\begin{align*}
L_{\uic}(\nWb;\vx) =& \left(H(V,s,t,x_1,x_2)\right)^2 + \lambda_2\left(V(s,T,x_1,x_2)\right)^2 + \left(V(s_{\lb},t,x_1,x_2)\right)^2 \\
&+ \left(V(\ln(B),t,x_1,x_2)-\vc(\ln(B),t,x_1,x_2)\right)^2, \\
L_{\uip}(\nWb;\vx) =& \left(H(V,s,t,x_1,x_2)\right)^2 + \lambda_2\left(V(s,T,x_1,x_2)\right)^2 + \left(V(s_{\lb},t,x_1,x_2)\right)^2 \\
&+ \left(V(\ln(B),t,x_1,x_2)-\vp(\ln(B),t,x_1,x_2)\right)^2, 
\end{align*} 
where $H(V,s,t,x_1,x_2)$ is defined in Equation \eqref{eq:pde_bergomi}. $\lambda_2$ is a constant with the default value $\lambda_2=25$, which strengthens the initial boundary condition. 

The loss functions for down-and-in options are defined as  
\begin{align*}
L_{\dic}(\nWb;\vx) =& \left(H(V,s,t,x_1,x_2)\right)^2 + \lambda_2\left(V(s,T,x_1,x_2)\right)^2 + \left(V(s_{\ub},t,x_1,x_2)\right)^2 \\
&+ \left(V(\ln(B),t,x_1,x_2)-\vc(\ln(B),t,x_1,x_2)\right)^2, \\
L_{\dip}(\nWb;\vx) =& \left(H(V,s,t,x_1,x_2)\right)^2 + \lambda_2\left(V(s,T,x_1,x_2)\right)^2 + \left(V(s_{\ub},t,x_1,x_2)\right)^2 \\
&+ \left(V(\ln(B),t,x_1,x_2)-\vp(\ln(B),t,x_1,x_2)\right)^2. 
\end{align*} 

The losses are defined on a single sample $\vx$. Given multiple samples $\vx_{(1)},\vx_{(2)},\dots,\vx_{(n)}$, the total loss is an average of the individual losses, i.e., 
\begin{align*}
	L_{\text{b,avg}}\left(\nWb;\{\vx_{(j)}\}_{j=1}^n\right)=\begin{cases}
		\frac{1}{n} \sum_{j=1}^{n} L_{\uic}(\nWb;\vx_{(j)}),&\text{for up-and-in call,} \\
		\frac{1}{n} \sum_{j=1}^{n} L_{\uip}(\nWb;\vx_{(j)}),&\text{for up-and-in put,}\\
		\frac{1}{n} \sum_{j=1}^{n} L_{\dic}(\nWb;\vx_{(j)}),&\text{for down-and-in call,}\\
		\frac{1}{n} \sum_{j=1}^{n} L_{\dip}(\nWb;\vx_{(j)}),&\text{for down-and-in put.}
	\end{cases} 
\end{align*}
We train the neural networks of vanilla options, get $\vc(\vx)$ and $\vp(\vx)$ and fix them before we train the networks of barrier options. After that, the loss function $L_{\text{b,avg}}$ is minimized only w.r.t. $\nWb$ and only the neural network of barrier options is trained.

\section{Numerical experiments}\label{sec:num_exp}
\subsection{Piecewise constant $\xi_0^t$}
In the Bergomi model, the model input $\xi_0^t$ is a function over $[0,T]$ and we consider the family of step functions
\begin{align*}
	\xi_0^t = \xi_j,\,\,\text{if}\,\,\,t_{j-1}\leq t\leq t_{j}
\end{align*}
given the nodes $0=t_0<t_1<\dots<t_m$, where $\{\xi_j\}_{j=1}^m$ are parameters. In the numerical experiments, we test the following two cases:
\begin{itemize}
	\item The constant case $\xi_0^t = \xi,\forall 0\leq t\leq 3$ as a baseline. In this case the network input is 
	\begin{align*}
	\vx=(s,t,x_1,x_2,T,B,r,q,\xi,\omega,k_1,k_2,\theta,\rho_1,\rho_2,\rho_{1,2}).
	\end{align*}
	\item The nine-segment case where $m=9$ and 
$$(t_j)_{j=1}^9=(1/52,1/26,1/12,1/6,1/4,1/2,1,2,3).$$
The nodes permit enough flexibility for options with both short and long time to maturities. In this case the network input is 
	\begin{align*}
	\vx=(s,t,x_1,x_2,T,B,r,q,\xi_1,\dots,\xi_9,\omega,k_1,k_2,\theta,\rho_1,\rho_2,\rho_{1,2}).
	\end{align*}
	The input dimension is 24 and the neural network is employed to deal with the high-dimensional case. 
\end{itemize} 

\subsection{Parameter range and sampling}\label{subsec:range}
Although the proposed method is unsupervised and does not need labels of prices generated from other pricing methods for training, we still need random samples for training. We also need option prices calculated from a benchmark method that are used to evaluate the results of neural networks after training. Here are the ranges of the parameters following the same constraints in both training and test samples.
\begin{align*}
	\begin{array}{cccc}
	K=100, & 0\leq T\leq 3, & 0.05^2\leq \xi_0^t\leq 0.5^2,\\
	0\leq r,q\leq 0.1, & 0\leq \omega \leq 3, & 0\leq \theta\leq 1,\\
	0.1\leq k_1\leq 4, & 2\leq k_2\leq 12, &  -0.9\leq \rho_1,\rho_2\leq 0.2.\\ 
\end{array}
\end{align*}
We choose a feasible range for each parameter. For example, $k_1$ and $k_2$ are chosen such that $X_t^{(1)}$ and $X_t^{(2)}$ are long-time and short-time volatility factors. $\rho_1$ and $\rho_2$ are mostly negative since returns and volatilities are usually negatively correlated. $\xi_0^t$ is similar to $\sigma^2$ in the BMS model and its range is chosen based on the scale of volatility.
The parameter $\rho_{1,2}$ needs to satisfy the following constraints to ensure the positive semidefinite property of the covariance matrix of the correlated Brownian motions 
$$\rho_1\rho_2-\sqrt{(1-\rho_1^2)(1-\rho_2^2)}\leq \rho_{1,2}\leq \rho_1\rho_2+\sqrt{(1-\rho_1^2)(1-\rho_2^2)}.$$
The variables $t$, $x_1$ and $x_2$ follow different constraints in training and test samples:
\begin{align*}
	\begin{array}{ccc}\hline
	 & \text{Training range} & \text{Test range}\\\hline
	t & 0\leq t\leq T & t=0 \\
	x_1 & x_{1,\lb}(\vx)\leq x_1 \leq x_{1,\ub}(\vx) & x_1=0\\
	x_2 & x_{2,\lb}(\vx)\leq x_2 \leq x_{2,\ub}(\vx) & x_2=0\\\hline
\end{array}
\end{align*}
where $$x_{j,\ub}(\vx)=-x_{j,\lb}(\vx)=3\sqrt{1/(2k_j)+0.01}$$ for $j=1,2$. The bound for $x_1$ and $x_2$ is built according to the variance of the limiting distribution of the OU process ${1}/{(2k_j)}$. These variables are equal to 0 in the test samples since we only need the option price at time $0$, i.e., $V(s,0,0,0)$. The ranges of $\ln(B)$ and $s$ are trickier since they are dependent on the option type. We sample $\ln(B)$ instead of $B$ based on the following rules
\begin{align*}
	\begin{array}{cc}\hline
	 & \text{Range for both training and test samples} \\\hline
	\text{vanilla options} & \text{(not applicable)} \\
	\text{up-and-in/out call} & \ln(K)\leq \ln(B)\leq \ln(1.5K)\\
	\text{down-and-in/out put} & \ln(K/1.5)\leq \ln(B)\leq \ln(K)\\
	\text{others} & \ln(K/1.5)\leq \ln(B)\leq \ln(1.5K)\\\hline
\end{array}
\end{align*}
The range of $\ln(B)$ is halved for up-and-in/out calls and down-and-in/out puts since we only calculate the non-degenerate case and the degenerate case falls into vanilla options.
Finally, the range of $s$ is listed for each case as follows:
\begin{align*}
	\begin{array}{ccc}\hline
	 & \text{Training range} & \text{Test range} \\\hline
	\text{vanilla options} & \ln(K/20)\leq s\leq \ln(20K) & \ln(K/2)\leq s\leq \ln(2K)\\
	\text{up-and-in/out option} & \ln(K/20)\leq s\leq \ln(B) & \ln(K/2)\leq s\leq \ln(B)\\
	\text{down-and-in/out option} & \ln(B)\leq s\leq \ln(20K) & \ln(B)\leq s\leq \ln(2K)\\\hline
\end{array}
\end{align*}
The range of $s$ is narrower in the test samples since we would like to focus more on the liquid options.

After choosing the range of each argument in $\vx$, we introduce how to sample them within the given range. All variables and parameters are sampled from the uniform distribution over the given intervals. If the lower and upper boundaries depend on other parameters, we use the conditional uniform distribution given the parameters in their boundaries. For example, $\rho_{1,2}$ follows the conditional uniform distribution over $$\left[\rho_1\rho_2-\sqrt{(1-\rho_1^2)(1-\rho_2^2)},\rho_1\rho_2+\sqrt{(1-\rho_1^2)(1-\rho_2^2)}\right]$$ given $\rho_1$ and $\rho_2$. The only exceptions are $x_1$ and $x_2$. The variables $(x_1,x_2)$ are sampled from their marginal distribution at time $t$, which is the two-dimensional normal distribution
\begin{align*}
	N\left(\left(\begin{array}{c}0\\0\\\end{array}\right),\left(\begin{array}{cc}
	\frac{1-e^{-2k_1 t}}{2k_1}+0.01 & \rho_{1,2}\frac{1-e^{-(k_1+k_2) t}}{k_1+k_2}\\
	\rho_{1,2}\frac{1-e^{-(k_1+k_2) t}}{k_1+k_2}& \frac{1-e^{-2k_2 t}}{2k_2}+0.01\\
\end{array}\right)\right)
\end{align*} and is then clipped within their range $x_{j,\lb}(\vx)\leq x_j \leq x_{j,\ub}(\vx),\forall j=1,2$. 0.01 is added to their variances to prevent the degenerate distribution at time $t=0$.
  
\subsection{Training and results}
We consider the neural network defined in Equation \eqref{eq:net_vanilla} consisting of $L=5$ layers for vanilla options and the neural network defined in Equation \eqref{eq:net_barrier} consisting of $(L_1,L_2)=(3,2)$ layers for barrier options. Each hidden layer contains $n=500$ neurons. The same network is used for the constant $\xi_0^t$ case and nine-segment $\xi_0^t$ case. The activation function $g$ is SiLU. The training batch size is 1000 and the training size is determined as follows:
\begin{align*}
	\begin{array}{ccc}\hline
	\text{Training size} & \text{Vanilla options} & \text{Barrier options} \\\hline	
	\text{constant}\,\xi_0^t & 10,000,000 & 20,000,000 \\
	\text{9-segment}\,\xi_0^t & 100,000,000 & 200,000,000 \\\hline
	\end{array}
\end{align*}
There are 10,000 test samples in each case. We use the Adam algorithm \cite{kingma2014adam} for training. The network is trained for 45 epochs in the constant $\xi_0^t$ case and 9 epochs in the nine-segment $\xi_0^t$ case. The learning rate decreases exponentially from $10^{-3}$ to $10^{-5}$.

We use simulation to calculate the benchmark, which is introduced in Appendix \ref{app:simulation_vanilla} and \ref{app:simulation_barrier}. The results are summarized in Table \ref{tab:rmse}. In Table \ref{tab:rmse}, we list the root mean square error (RMSE) of the neural network solution for each option, i.e., 
\begin{align*}
	\text{RMSE} = \sqrt{\frac{1}{n}\sum_{1\leq j\leq n}\left(V(\vx_{(j)})-\bar{V}(\vx_{(j)})\right)^2},
\end{align*}
where $\vx_{(j)}$ are the test samples following the ranges and distributions in Section \ref{subsec:range}, $V(\vx_{(j)})$ is the solution given by the neural network and $\bar V(\vx_{(j)})$ is the benchmark. The RMSE is calculated over the 10,000 test samples, which have not been used during the training process. We also list the RMSE of the benchmark
\begin{align*}
	\sqrt{\frac{1}{n}\sum_{1\leq j\leq n}\left(\text{se}(\bar{V}(\vx_{(j)}))\right)^2},
\end{align*}
where $\text{se}(\bar{V}(\vx_{(j)}))$ is the standard error of the estimate $\bar{V}(\vx_{(j)})$ in the simulation benchmark. Since the benchmarks are noisy, the RMSE of the neural network cannot be much smaller than the RMSE of the benchmark.

\begin{table}[h]
\centering
\begin{tabular}{lcccc}\hline
& \multicolumn{2}{c}{RMSE of network solutions} & \multicolumn{2}{c}{RMSE of simulation} \\ 
Option & constant $\xi_0^t$ & 9-segment $\xi_0^t$ & constant $\xi_0^t$ & 9-segment $\xi_0^t$\\\hline
vanilla call & 0.0686 & 0.0685 & 0.1422 & 0.1415 \\
vanilla put & 0.1066 & 0.1039 & 0.0926 & 0.0940 \\
up-and-out call & 0.0772 & 0.1031 & 0.0622 & 0.0640 \\
up-and-in call & 0.1117 & 0.1309 & 0.1225 & 0.1146 \\
down-and-out call & 0.1479 & 0.1676 & 0.2060 & 0.2094 \\
down-and-in call & 0.1329 & 0.1576 & 0.1336 & 0.1427 \\
up-and-out put & 0.1133 & 0.1372 & 0.1063 & 0.1067 \\
up-and-in put & 0.1069 & 0.1336 & 0.0963 & 0.1000 \\
down-and-out put & 0.0736 & 0.0923 & 0.0600 & 0.0617 \\
down-and-in put & 0.1171 & 0.1271 & 0.0992 & 0.0999 \\
\hline 
\end{tabular}
\caption{The root mean square error (RMSE) of the neural network solution over the 10,000 test samples for each option, compared with the RMSE of the simulation benchmark.}
\label{tab:rmse}
\end{table}

\subsection{Fitted curves}
\begin{figure}[h]
\centering
	\includegraphics[width=1\textwidth]{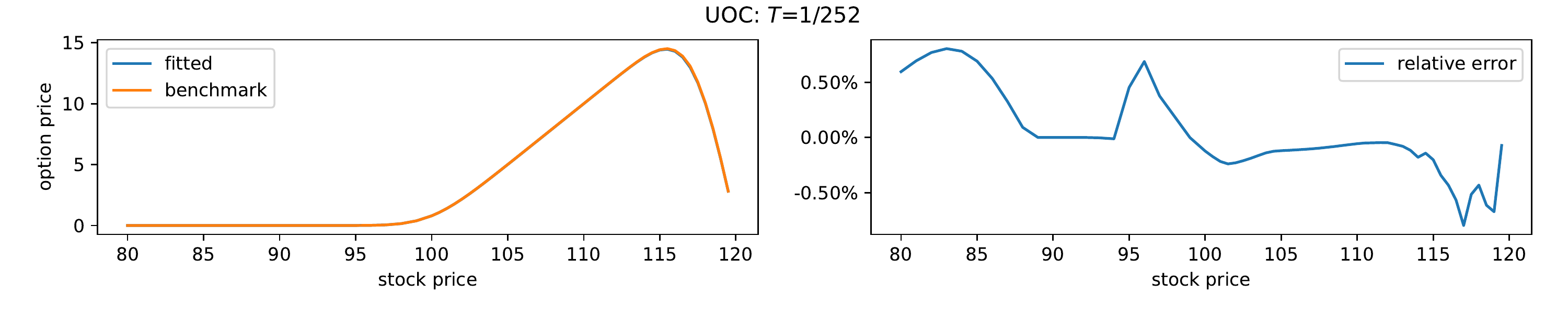}
	\includegraphics[width=1\textwidth]{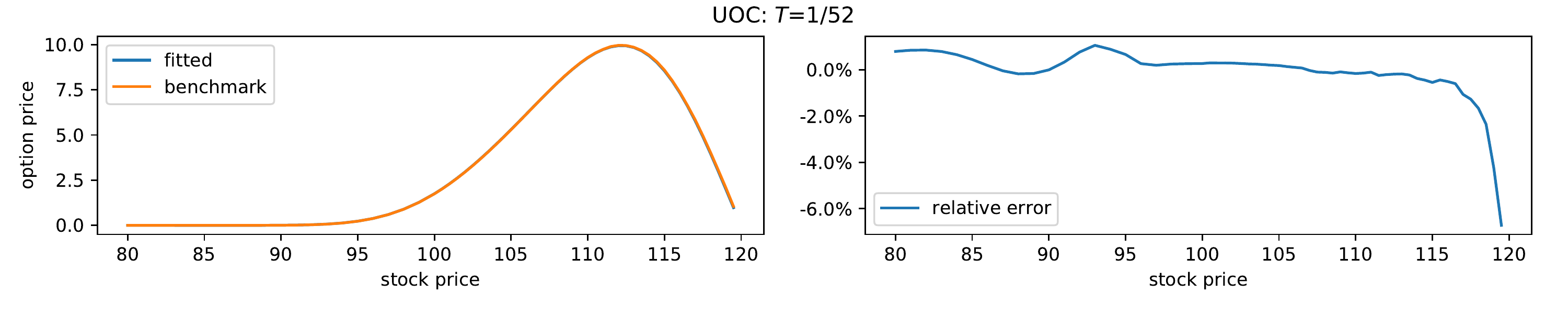}
	\includegraphics[width=1\textwidth]{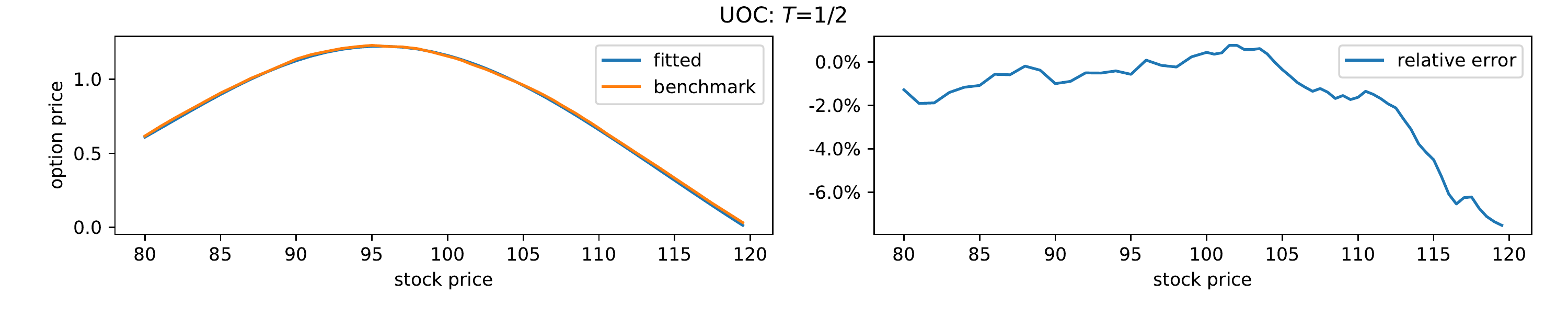}
	\includegraphics[width=1\textwidth]{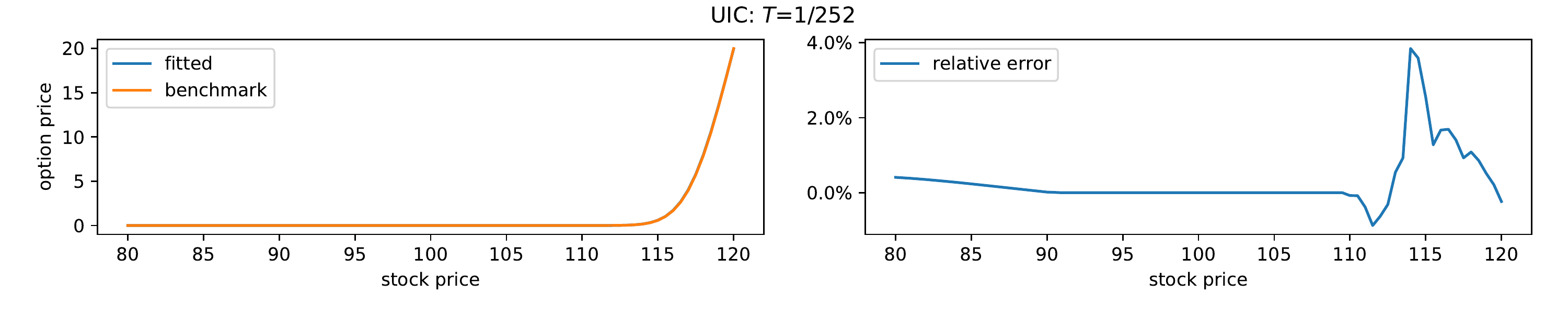}
	\includegraphics[width=1\textwidth]{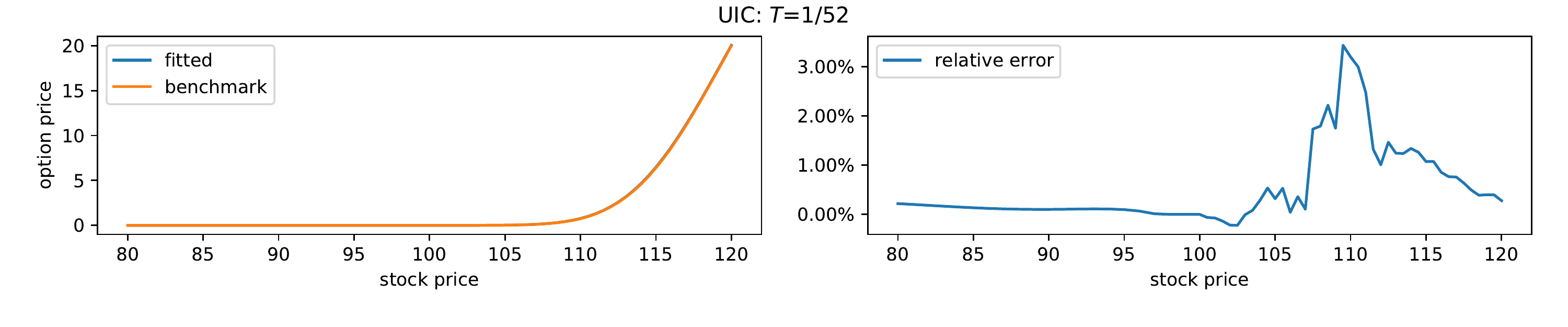}
	\includegraphics[width=1\textwidth]{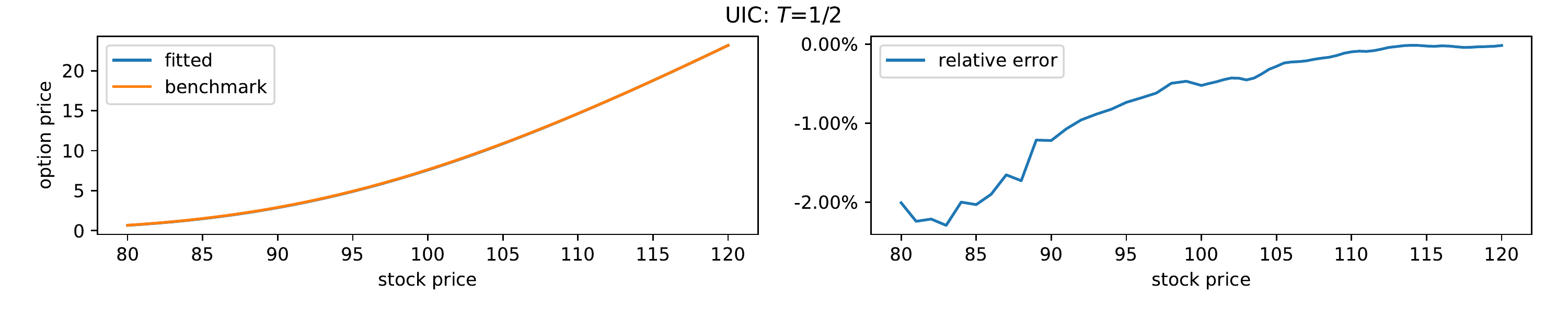}
\caption{Comparison of the fitted neural network solution and the simulation benchmark of the up-and-out/in call when $K=100,B=120,r=q=0,\xi_0^t=0.1,\omega=1,k_1=1,k_2=10,\theta=0.5,\rho_1=\rho_2=-0.5,\rho_{1,2}=0$ and $t=x_1=x_2=0$. UOC and UIC stand for the up-and-out and up-and-in call respectively.}
\label{fig:fitting_uoc}
\end{figure}

\begin{figure}[h]
\centering
	\includegraphics[width=1\textwidth]{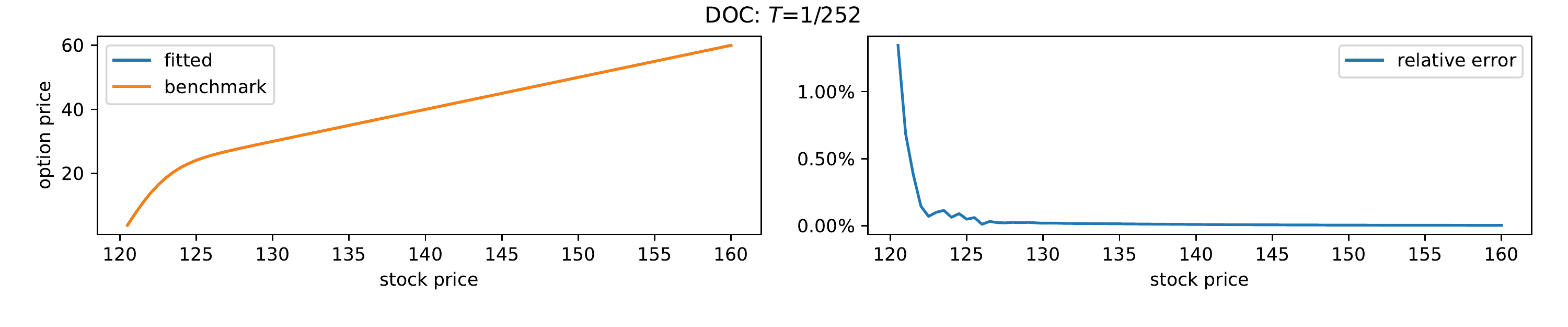}
	\includegraphics[width=1\textwidth]{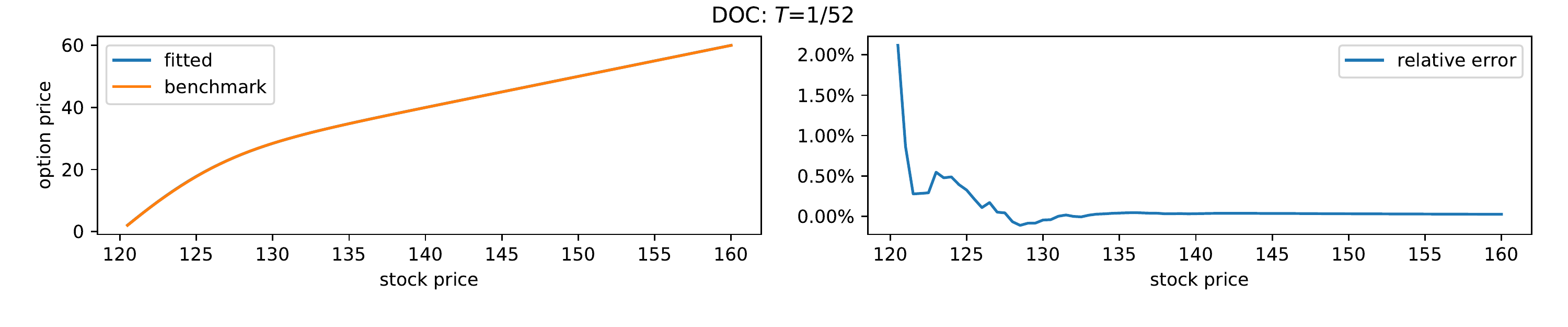}
	\includegraphics[width=1\textwidth]{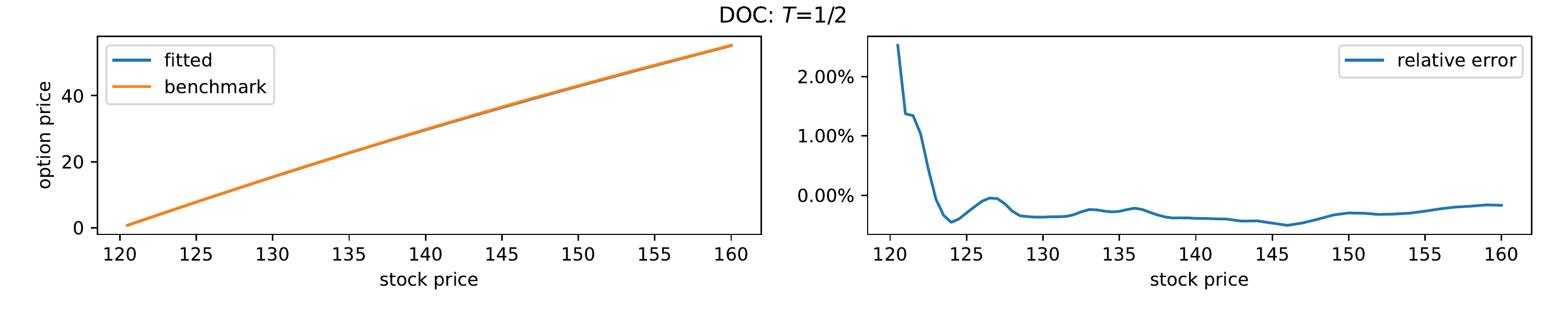}
	\includegraphics[width=1\textwidth]{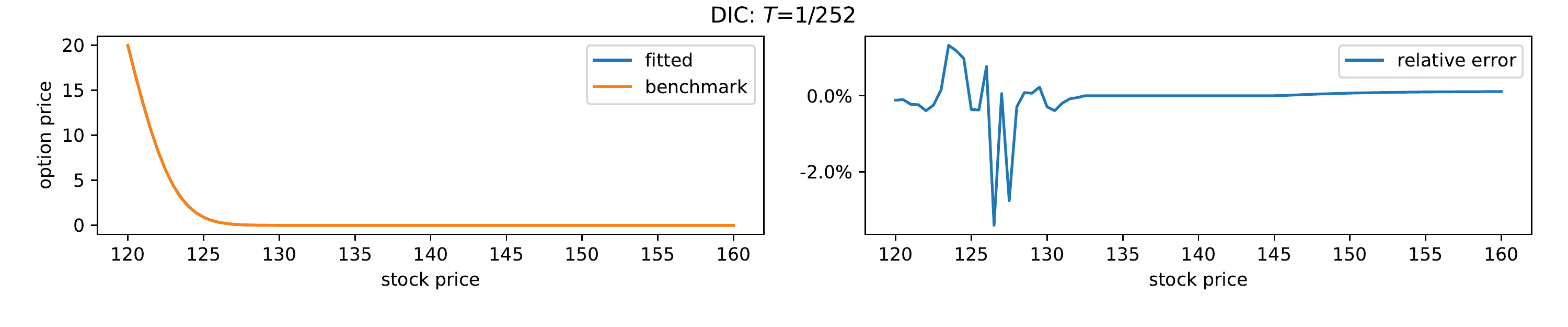}
	\includegraphics[width=1\textwidth]{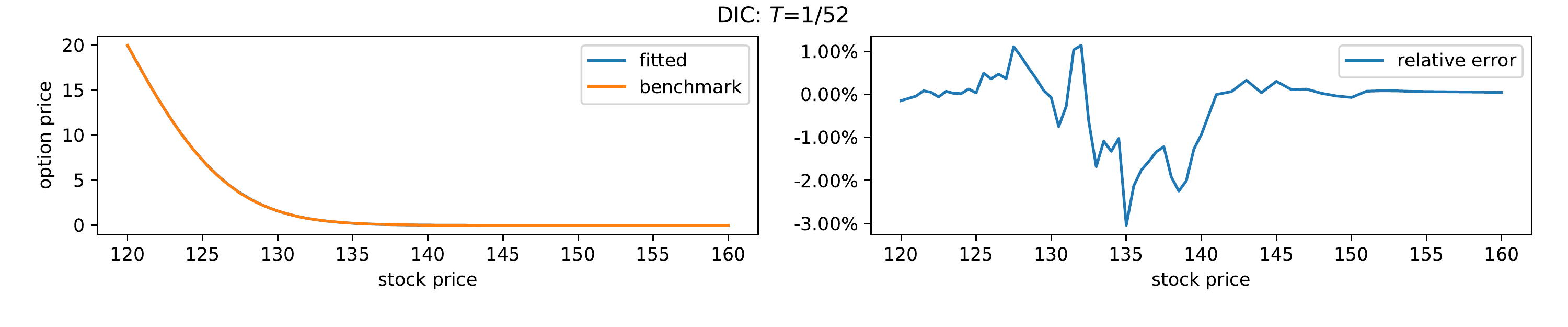}
	\includegraphics[width=1\textwidth]{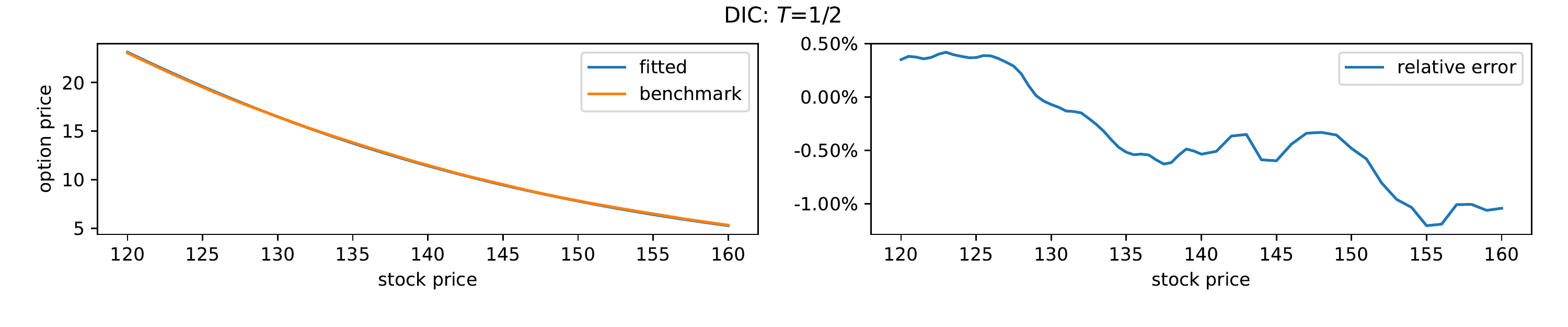}
\caption{Comparison of the fitted neural network solution and the simulation benchmark of the down-and-out/in call when $K=100,B=120,r=q=0,\xi_0^t=0.1,\omega=1,k_1=1,k_2=10,\theta=0.5,\rho_1=\rho_2=-0.5,\rho_{1,2}=0$ and $t=x_1=x_2=0$. DOC and DIC stand for the down-and-out and down-and-in call respectively.}
\label{fig:fitting_doc}
\end{figure}

The singular terms are included in the neural networks such that the non-smooth and discontinuous boundary conditions can be fitted. Thus the neural networks keep the singular properties around $(s,t)=(\ln(K),T)$ and $(s,t)=(\ln(B),T)$ and are good at fitting option price curves of short maturities. Consequently, they are also able to replicate the prices of longer maturity given they are fitted to satisfy the PDE.
In Figures \ref{fig:fitting_uoc} and \ref{fig:fitting_doc}, we show the fitted neural network solution and the simulation benchmark of the barrier calls at $T=1/252$ (1 day), $T=1/52$ (1 week) and $T=1/2$ (half a year) as well as the relative error $(V(\vx)-\bar V(\vx))/\max(\bar V(\vx),h)$ as an example. For these examples we use $h=0.25$. The barrier calls are taken as the examples since the barrier puts are bounded and are usually fitted with smaller errors.

\subsection{Calculation speed}
The neural network can calculate prices of a batch of parameter sets at the same time. Thus it will be super fast to generate the option prices once trained. The calculation times of the neural networks used in the numerical experiments are summarized in Table \ref{tab:compute_time}. Note that the vanilla and knock-in options only use one neural network, while the knock-out options are calculated as a difference of the vanilla and knock-in options and make use of two networks. By means of the GPU acceleration, 200,000 prices can be calculated in 0.133 seconds as most.
\begin{table}[h]
\centering
\begin{tabular}{llccccccc}\hline
 & Input size &  1 & 10 & 100 & 1k & 10k & 100k & 200k \\\hline
vanilla \& & GPU(s) & 0.017 & 0.027 & 0.021 & 0.021 & 0.025 & 0.055 & 0.070 \\
knock-in & CPU(s) & 0.032 & 0.033 & 0.034 & 0.092 & 0.533 & 4.943 & 9.296 \\
 \hline
\multirow{2}{*}{knock-out} & GPU(s) & 0.070 & 0.062 & 0.067 & 0.062 & 0.070 & 0.083 & 0.133 \\ 
 & CPU(s) & 0.078 & 0.073 & 0.093 & 0.194 & 1.347 & 10.717 & 18.585 \\
 \hline
\end{tabular}
	\caption{Computation time of the neural network solutions in the numerical experiments consisting of 5 layers, with 500 neurons in each layer. CPU is an Intel Xeon CPU @ 2.20GHz. GPU is a Tesla V100-SXM2-16GB. All times are in seconds.}
	\label{tab:compute_time}
\end{table}
                 
\section{Conclusion}\label{sec:conclusion}
In this paper, we have developed an unsupervised deep learning method to solve the barrier options under the two-factor Bergomi model. The neural networks serve as the approximate option surfaces and are trained to satisfy the PDE as well as the boundary conditions. A trained neural network can calculate option values extremely fast. 
 
Here we summarize the main innovations based on the unsupervised deep learning method:
\begin{itemize}
\item We propose two singular terms to deal with the non-smoothness at the strike level and the discontinuity at the barrier level so that the neural network can fit the boundary conditions of the barrier options.
\item We use six networks to express the eight barrier options in one framework. We do not train the eight options separately, but make use of the in-out parity. We build networks for knock-in options given they contain only one singularity and are easier to be fitted.  
\item The neural network is employed to deal with the high dimensionality coming with the large number of parameters and the function input in the multi-factor forward variances in the Bergomi model.
\item Boundary conditions of the volatility factors are estimated by the BMS model, which increases the accuracy of the method. 
\end{itemize} 

The proposed method can also deal with the other stochastic volatility models, as long as we find the suitable boundary conditions of volatility and the suitable estimate. The other stochastic volatility models should not be more complex than the Bergomi model given there are fewer parameters and there is no function input in the model. So the method for the Bergomi model serves as a good example of the applications to the stochastic volatility models.
 
 The two proposed singular terms are good examples for the case that we need to solve heat equations with non-smooth or discontinuous initial conditions. We can incorporate multiple singular terms into one neural network for more complex initial conditions, which would facilitate fitting the asymptotic behaviors of the solution near the initial condition. Moreover, the idea of singular terms can be extended to deal with other types of problems as long as we know the overall shape and approximate position of the non-smoothness or discontinuity in their solutions.

\clearpage
\section*{}
\begin{center}
\Large\textbf{Acknowledgement} 
\end{center}
We are grateful to Alireza Javaheri and Mehdi H. Sonthonnax of Credit Suisse for their time and expertise in the Bergomi model, which is indispensable for the numerical experiments.
  
\bibliographystyle{abbrv}
\bibliography{references.bib}

\clearpage
\appendix
\section{PDE for the Bergomi model}\label{app:pde}
The PDE \eqref{eq:pde_bergomi} for the two-factor Bergomi model is derived according to the multidimensional version of the Feynman-Kac formula (see Theorem 1.3.17 in \cite{pham2009continuous}). In Section \ref{subsec:bergomi}, we introduce the dynamics of the Bergomi model, which are summarized as
\begin{align*}
	\dd{s}_t &=(r-q-\sigma^2(t,X^{(1)}_t, X^{(2)}_t)/2)\dt + \sigma(t,X^{(1)}_t, X^{(2)}_t)\dW_t^{(S)},\\	
	\dd{X}_t^{(1)}&=-k_1 X_t^{(1)}\dt+\dW_t^{(1)},\\
	\dd{X}_t^{(2)}&=-k_2 X_t^{(2)}\dt+\dW_t^{(2)},
\end{align*}
where $s_t=\ln(S_t)$ is the log-price process and $\sigma(t,x_1, x_2)$ satisfies $\sigma^2(t,X^{(1)}_t, X^{(2)}_t)=\xi_{t}^{t}$. The correlations of the Brownian motions are $\dW_t^{(S)}\dW_t^{(i)}=\rho_{i}\dt,\forall i=1,2$ and $\dW_t^{(1)}\dW_t^{(2)}=\rho_{1,2}\dt$. For an applicable function $V(s,t,x_1,x_2)$, the infinitesimal generator $\mathcal{L}_t$ is defined as
\begin{align*}
\mathcal{L}_t V=&(r-q-\sigma^2(t,x_1,x_2)/2)\frac{\partial V}{\partial s} -k_1 x_1\frac{\partial V}{\partial x_1}-k_2 x_2\frac{\partial V}{\partial x_2} \\
	&+\frac{1}{2}\sigma^2(t,x_1,x_2)\frac{\partial^2 V}{\partial s^2}+\frac{1}{2}\frac{\partial^2 V}{\partial x_1^2}+\frac{1}{2}\frac{\partial^2 V}{\partial x_2^2}\\
	& + \rho_1 \sigma(t,x_1,x_2)\frac{\partial^2 V}{\partial s \partial x_1} + \rho_2 \sigma(t,x_1,x_2)\frac{\partial^2 V}{\partial s \partial x_2} + \rho_{1,2}\frac{\partial^2 V}{\partial x_1 \partial x_2} .	
\end{align*} 
According to the Feynman-Kac formula, the vanilla options 
\begin{align*}
	V(s,t,x_1,x_2)=\E\left(e^{-r(T-t)}(\eta(e^{s_{T}}-K))^+\,\vert S_t=e^{s},X_t^{(1)}=x_1,X_t^{(2)}=x_2\right)
\end{align*}
satisfy the equation 
\begin{align*}
\frac{\partial V}{\partial t}+\mathcal{L}_t V-rV=0.
\end{align*}

The barrier options are path-dependent and cannot be fully explained by the Feynman-Kac formula. However, they also satisfy the equation and this can be explained by the no-arbitrage property of the option values: the discounted option value should be a martingale. The increment of the discounted option value is 
\begin{align*}
	\dd{(e^{-rt}V(s_t,t,{X}_t^{(1)},{X}_t^{(2)}))}=&-re^{-rt}V\dt + e^{-rt}\frac{\partial V}{\partial t}\dt+e^{-rt}\mathcal{L}_t V\dt\\
	&+e^{-rt}\frac{\partial V}{\partial s}\sigma\dW_t^{(S)} + e^{-rt}\frac{\partial V}{\partial x_1}\dW_t^{(1)} + e^{-rt}\frac{\partial V}{\partial x_2}\dW_t^{(2)}.
\end{align*} 
The drift term of the increment of a martingale should be 0, which leads to the same equation.

\section{Black-Scholes formula of vanilla and barrier options}\label{app:bs_formula}
The BS formula of vanilla options was proposed in \cite{black_scholes_1973}. We use the variant with the dividend rate. Suppose $s$ is the log-price, $K$ is the strike, $t$ is the current time, $T$ is the maturity (expiration) time, $r$ is the risk-free interest rate and $q$ is the dividend rate. The vanilla call and put are priced using
\begin{align*}
\vc(s;K) &=  e^{s-q(T-t)}N(h/v+  v/2)
- K e^{-r(T-t)}N(h/v - v/2),\\
\vp(s;K) &= -e^{s-q(T-t)}N(-( h/v+  v/2))
+ K e^{-r(T-t)}N(- (h/v - v/2)),
\end{align*}
where
\begin{align*}
h &= s - \ln(K) + (r-q)(T-t),\\
v &= \sigma\sqrt{T-t}.
\end{align*}

The formula of barrier options needs the digital call and put of which the payoffs are $\1{S_T>K}$ and $\1{S_T<K}$ and the prices are
\begin{align*}
\cd(s;K) &=  e^{-r(T-t)}N(h/v - v/2),\\
\pd(s;K) &= e^{-r(T-t)}N(-h/v + v/2).
\end{align*}

Suppose $B$ is the barrier level, we summarize the pricing formulae in \cite{barrier_formula} as follows:
\begin{align*}
\uic(s;K,B)=&\vc(s;B)+(B-K)\cd(s;B)\\
&+\delta(\vc(\tilde{s};K)-\vc(\tilde{s};B)+(K-B)\cd(\tilde{s};B)),\\
\uoc(s;K,B)=&\vc(s;K)-\uic(s;K,B),\\
\dic(s;K,B)=&\vc(s;K)-\vc(s;\max(B,K))-\max(0,B-K)\cd(s;B)\\
&+\delta(\vc(\tilde{s};\max(B,K))+\max(0,B-K)\cd(\tilde{s};B)),\\
\doc(s;K,B)=&\vc(s;K)-\dic(s;K,B),\\
\uip(s;K,B)=&\vp(s;K)-\vp(s;\min(B,K))-\max(0,K-B)\pd(s;B)\\
&+\delta(\vp(\tilde{s};\min(B,K))+\max(0,K-B)\pd(\tilde{s};B)),\\
\uop(s;K,B)=&\vp(s;K)-\uip(s;K,B),\\
\dip(s;K,B)=&\vp(s;B)-(B-K)\pd(s;B)\\
&+\delta(\vp(\tilde{s};K)-\vp(\tilde{s};B)+(B-K)\pd(\tilde{s};B)),\\
\dop(s;K,B)=&\vp(s;K)-\dip(s;K,B),\\
\delta=&(e^s/B)^{1+(2(q-r))/\sigma^2},\\
\tilde{s}=&2\ln(B)-s,
\end{align*}
where the up-and-in/out formulae are applicable where $s\leq\ln(B)$ and the down-and-in/out formulae are applicable where $s\geq\ln(B)$. Furthermore, the up-and-in/out calls are applicable when $B\geq K$ and the down-and-in/out puts are applicable when $B\leq K$.

\section{Benchmark of vanilla options}\label{app:simulation_vanilla}
The dynamics of the Bergomi model in Section \ref{subsec:bergomi} are summarized as
\begin{align}
\begin{split}
	\dd{s}_t &=(r-q-\xi_t^t/2)\dt + \sqrt{\xi_t^t}\dW_t^{(S)},\\
	\xi_{t}^{t}&=\xi_{0}^{t}\exp\left(\omega x_t^t-\frac{\omega^2}{2}\text{var}(x_t^t)\right),\\
	x_t^t&=\alpha_{\theta}\left((1-\theta)X_t^{(1)}+\theta X_t^{(2)}\right),\\	
	\dd{X}_t^{(1)}&=-k_1 X_t^{(1)}\dt+\dW_t^{(1)},\\
	\dd{X}_t^{(2)}&=-k_2 X_t^{(2)}\dt+\dW_t^{(2)},
\end{split}\label{eq:bergomi_dynamic}
\end{align}
where $s_t=\ln(S_t)$. The correlated Brownian motions can be expressed using independent Brownian motions $Z_t^{(j)},j=1,2,3$ as
\begin{align*}
	W_t^{(1)} &=Z_t^{(1)},\\
	W_t^{(2)} &=\mu_{21}Z_t^{(1)}+\mu_{22}Z_t^{(2)},\\
	W_t^{(S)} &=\mu_{31}Z_t^{(1)}+\mu_{32}Z_t^{(2)}+\mu_{33}Z_t^{(3)},
\end{align*} 
where $\mu_{21}=\rho_{1,2}$, $\mu_{22}=\sqrt{1-\rho_{1,2}^2}$, $\mu_{31}=\rho_{1}$, $\mu_{32}=\frac{\rho_2-\rho_1\rho_{1,2}}{\sqrt{1-\rho_{1,2}^2}}$ and $$\mu_{33}=\sqrt{\frac{1-\rho_1^2-\rho_2^2-\rho_{1,2}^2+2\rho_1\rho_2\rho_{1,2}}{1-\rho_{1,2}^2}}.$$

Clearly $s_t$ is dependent on $X_t^{(1)}$ and $X_t^{(2)}$ but not conversely. Thus we can determine the volatility process first and then the stock price process. This means the vanilla option prices can be evaluated given the condition of volatility. For example, the call option is 
\begin{align*}
	\E((S_T-K)^+\vert S_{0})=\left.\E\left(\E\left((S_T-K)^+\left| \{X_t^{(1)}\}_{t=0}^{T},\{X_t^{(2)}\}_{t=0}^{T}, S_{0}\right)\right.\right|S_{0}\right).
\end{align*}
Given $\{X_t^{(1)}\}_{t=0}^{T}$ and $\{X_t^{(2)}\}_{t=0}^{T}$, we also know the paths of $\{Z_t^{(1)}\}_{t=0}^{T}$, $\{Z_t^{(2)}\}_{t=0}^{T}$ and $\{\xi_t^t\}_{t=0}^{T}$. The SDE of $s_t$ becomes 
\begin{align*}
	\dd{s}_t =&(r-q-\xi_t^t/2)\dt + \sqrt{\xi_t^t}(\mu_{31}\dd{Z}_t^{(1)}+\mu_{32}\dd{Z}_t^{(2)}+\mu_{33}\dd{Z}_t^{(3)})\\
	=&(r-q-(\mu_{31}^2+\mu_{32}^2)\xi_t^t/2)\dt + \sqrt{\xi_t^t}(\mu_{31}\dd{Z}_t^{(1)}+\mu_{32}\dd{Z}_t^{(2)})\\&-\mu_{33}^2\xi_t^t/2\,\dt+\sqrt{\xi_t^t}\mu_{33}\dd{Z}_t^{(3)}
\end{align*}
where only $Z_t^{(3)}$ is random and the volatility function is fixed. The equivalent spot is  
\begin{align*}
	\widetilde{S_{0}} = S_{0}\exp\left(\int_{0}^T\sqrt{\xi_t^t}(\mu_{31}\dd{Z}_t^{(1)}+\mu_{32}\dd{Z}_t^{(2)})-(\mu_{31}^2+\mu_{32}^2) \xi_{t}^{t}/2\, dt\right)
\end{align*}
and the equivalent volatility rate during $[0,T]$ is 
\begin{align*}
	\tilde\sigma_{0}^T = \sqrt{\frac{\mu_{33}^2}{T}\int_{0}^T \xi_{t}^{t}dt}
\end{align*}
The conditional expectation can be calculated by the Black-Scholes formula:
\begin{align*}
	\E((S_T-K)^+\vert \{X_t^{(1)}\}_{t=0}^{T},\{X_t^{(2)}\}_{t=0}^{T}, S_{0}) = \text{BS-}\vc(\widetilde{S_{0}},K,T,\tilde\sigma_{t}^T,r,q)
\end{align*} 
Then we just need to sample paths of $\{X_t^{(1)}\}_{t=0}^{T}$ and $\{X_t^{(2)}\}_{t=0}^{T}$ and take the average of the conditional expectation to get the vanilla option price. The same applies to the vanilla put. The variance of the conditional expectation is far less than the variance of trivial simulation. However, this approach does not work for the barrier options. Since the payoff of barrier options are path-dependent and we cannot get the equivalent spot and volatility rate. 

\section{Benchmark of barrier options}\label{app:simulation_barrier}
Since we cannot use conditional expectation for barrier options as in Appendix \ref{app:simulation_vanilla}, we need to sample the log-price $\{s_t\}_{t=0}^T$ for simulation. Note that $\xi_t^t$ contains an exponential function and could be very large. Under this case, $s_t$ converges to $-\infty$ quickly, and $S_t$ converges to 0 quickly. When we evaluate the barrier puts, this not a problem. However, this is a problem for barrier calls. There will be very few or no samples of positive values, and the barrier calls will be underestimate. As a result, the Euler scheme is directly applied to Equation \eqref{eq:bergomi_dynamic} to price barrier puts, while we use importance sampling to price barrier calls for variance reduction. 

The importance sampling is implemented according to Girsanov theorem \cite{girsanov}. First, let 
\begin{align*}
	\dd{{Z}}_t^{(3)} = \dd{\tilde{Z}}_t^{(3)} + \mu_{33}\sqrt{\xi_t^t}\dt
\end{align*}
where $\tilde{Z}_t^{(3)}$ is a Brownian motion under the measure $\mathbb{Q}$ while ${Z}_t^{(3)}$ is a Brownian motion under the measure $\mathbb{P}$ with the Radon-Nikodym derivative 
\begin{align*}
	\frac{\dd{\mathbb{P}}}{\dd{\mathbb{Q}}}=\exp\left(-\int_0^t \mu_{33}^2 \xi_{u}^{u}/2\, du- \int_0^t\mu_{33}\sqrt{\xi_{u}^{u}}\dd{\tilde{Z}}_u^{(3)}\right).
\end{align*}
After that we replace $\dd{{Z}}_t^{(3)}$ using $\dd{\tilde{Z}}_t^{(3)}$ in the SDE of $s_t$ such that 
\begin{align*}
	\dd{s}_t =&(r-q-(\mu_{31}^2+\mu_{32}^2)\xi_t^t/2)\dt + \sqrt{\xi_t^t}(\mu_{31}\dd{Z}_t^{(1)}+\mu_{32}\dd{Z}_t^{(2)})\\&-\mu_{33}^2\xi_t^t/2\,\dt+\sqrt{\xi_t^t}\mu_{33}\dd{Z}_t^{(3)}\\
	=&(r-q-(\mu_{31}^2+\mu_{32}^2)\xi_t^t/2)\dt + \sqrt{\xi_t^t}(\mu_{31}\dd{Z}_t^{(1)}+\mu_{32}\dd{Z}_t^{(2)})\\&+\mu_{33}^2\xi_t^t/2\,\dt+\sqrt{\xi_t^t}\mu_{33}\dd{\tilde{Z}}_t^{(3)}.
\end{align*}
We sample $\{s_t\}_{t=0}^{T}$ under $\mathbb{Q}$, i.e., 
\begin{align*}
	s_t = s_0+\int_0^t (r-q-(1-2\mu_{33}^2)\xi_u^u/2)\dd{u} + \sqrt{\xi_u^u}(\mu_{31}\dd{Z}_u^{(1)}+\mu_{32}\dd{Z}_u^{(2)}+\dd{\tilde{Z}}_u^{(3)}).
\end{align*}
Each sample path $\{s_t\}_{t=t_0}^{T}$ is attached the following weight
\begin{align*}
	\exp\left(-\int_0^T \mu_{33}^2 \xi_{u}^{u}/2\, du- \int_0^t\mu_{33}\sqrt{\xi_{u}^{u}}\dd{\tilde{Z}}_u^{(3)}\right).
\end{align*}
Since the drift term $-\mu_{33}^2\xi_t^t/2\,\dt$ in the original SDE is changed to $\mu_{33}^2\xi_t^t/2\,\dt$ in the SDE under $\mathbb{Q}$, there will be enough large samples of $s_T$ and the barrier call options will not be underestimated. After we collect enough sample paths, we use the definition of barrier options in Table \ref{tab:option_def} to evaluate them.

\end{document}